\documentclass[journal,hidelinks]{IEEEtran}
\usepackage{url,array,lipsum,hhline,graphicx,subfigure,amsfonts,cite,epstopdf,makeidx,outlines,textcomp,multirow,soul,hyperref,enumitem,amssymb,amsmath,color,graphicx, amsbsy,comment}
\usepackage[none]{hyphenat}
\allowdisplaybreaks
\usepackage{makecell}

\newcolumntype{M}[1]{>{\centering\arraybackslash}m{#1}}

\begin{document}
\title{Quantifying the Risk of Wildfire Ignition by Power Lines under Extreme Weather Conditions}

\author{Reza  Bayani,~\IEEEmembership{Graduate Student Member,~IEEE}, Muhammad Waseem,
~\IEEEmembership{Student Member,~IEEE}, Saeed D. Manshadi,~\IEEEmembership{Member,~IEEE}, and Hassan Davani
\thanks{ Reza Bayani is with University of California San Diego, La Jolla, CA, 92093, USA, and San Diego State University (email:rbayani@ucsd.edu). M. Waseem and S. D. Manshadi are with Department of Electrical and Computer Engineering, San Diego State University, San Diego, CA, 92182, USA (email:\{mwaseem2282;smanshadi\}@sdsu.edu). Hassan Davani is with Department of Civil, Construction, and Environmental Engineering, San Diego State University, San Diego, CA, 92182, USA (email:hdavani@sdsu.edu).
        }
}

\markboth{}%
{Shell \MakeLowercase{\textit{et al.}}: Bare Demo of IEEEtran.cls for IEEE Journals}

\maketitle

{
\begin{abstract}
Utilities in California conduct Public Safety Power Shut-offs (PSPSs) to eliminate the elevated chances of wildfire ignitions caused by power lines during extreme weather conditions. We propose Wildﬁre Risk Aware operation planning Problem (WRAP), which enables system operators to pinpoint the segments of the network that should be de-energized. Sustained wind and wind gust can lead to conductor clashing, which could ignite surrounding vegetation. The 3D non-linear vibration equations of power lines are employed to generate a dataset that considers physical, structural, and meteorological parameters. With the help of machine learning techniques, a surrogate model is obtained which quantifies the risk of wildfire ignition by individual power lines under extreme weather conditions. The cases illustrate the superior performance of WRAP under extreme weather conditions in mitigating wildfire risk and serving customers compared to the naive PSPS approach and another method in the literature. Cases are also designated to sensitivity analysis of WRAP to critical load-serving control parameters in different weather conditions. Finally, a discussion is provided to explore our wildfire risk monetization approach and its implications for WRAP decisions. 
\end{abstract}
}

\IEEEpeerreviewmaketitle
\vspace{-0.30cm}
\section{Introduction}
\vspace{-0.20cm}
\subsection{Motivation}
\bstctlcite{IEEEexample:BSTcontrol}

\IEEEPARstart{W}{ildfires} are among {the} threats that could potentially lead to disastrous events worldwide. Arguably, fire seasons are getting {longer} and more frequent, primarily due to climate change and global warming \cite{jolly2015climate}. The wildfire frequency and the burn{ed} area in the western U.S. have grown exponentially since 1950 \cite{weber2020spatiotemporal}. The causes of wildfire ignition include natural sources (lightning), human activities (arson, campfire, etc.), equipment faults, and power lines, among  other things \cite{syphard2015location}. Pacific Gas \& Electric Company (PG\&E) declared 414 fire ignition events during 2015-2017 caused by electric power lines \cite{pge_report}. According to the historical data during the period 1960-2009 in Southern California, fires ignited by power lines burn on average ten times the area burnt as a result of fires initiated by other sources \cite{mitchell2013power}. Extreme winds, such as the ones occurring during {the} fall season in Southern California, not only increase the risk of power line-related ignitions but also facilitate the propagation of fire. Reportedly, wildfires caused by power lines account for almost half of the most destructive fires in California history \cite{psps}. Conductor clashing (phase to phase faults), fall of the line on the ground (phase to ground faults), arcs, and contact with the surrounding vegetation are among the incidents related to power lines that could cause ignitions \cite{kandanaarachchi2020early}. 

\par In addition to the wind, temperature, humidity, and vegetation are among the meteorological and geographical factors correlated with the chances of fire ignitions by power lines \cite{sotolongo2020}, with {the} wind as the leading factor. Although low wind speeds will cool off conductors and can benefit dynamic line rating \cite{gentle2012concurrent}, high wind speeds can lead to faults within power lines. According to the analysis of the 11-year outage record data provided by San Diego Gas \& Electric (SDG\&E), for every 25 km/h increase in the wind gust speed, the outage probability is increased ten times \cite{mitchell2013power}. {That is one reason why higher wind speeds are associated with higher chances of wildfire ignition.}

\par High sustained wind speeds (average wind speed lasting over 10 minutes) typically occur during California's late summer and fall seasons. Wind gust{s} could significantly affect the swing angle of the conductors \cite{lihong2006parameters}. The combination of high sustained wind speeds and wind gusts could lead to conductor clashing, which, together with hot and dry weather conditions, {dramatically} increases the risk of wildfires taking place. The clashing of the energized power line conductors may cause vaporization and melting of the conductor materials and ejection of the molten metal as small particles \cite{sutlovic2019analysis}. Drop of hot particles due to conductor clashing on dried vegetation or direct contact between conductors and vegetation can potentially ignite wildfires \cite{jazebi2019review}. That is why utilities across California adopted \emph{Public Safety Power Shut-offs} (PSPS) to avoid such ignitions under extreme conditions with wind gusts exceeding 22.8 m/s.

\par {The} California Public Utilities Commission allows utilities to deliberately cut off power to electrical lines as a preventive measure in high-risk situations with a looming threat of wildfire ignition \cite{psps}. From 2013 to 2020, PG\&E, SDG\&E, Southern California Edison, and PacifiCorp collectively performed 51 PSPS events in multiple locations within California, which impacted 3.2 million customers \cite{murphy2021}. However, a PSPS event has its {own} risks and unfavorable consequences, particularly for medically vulnerable populations and low-income communities \cite{sotolongo2020}. In October 2019, power shut-offs by PG\&E affected 1.8 million customers, with outages lasting more than five days in some cases. PSPS events cost the California economy in {the} order of billion dollars \cite{cnbc2019}, while {a} lack of power supply could lead to casualties \cite{fox2019}. It can be concluded that although PSPS schedules are a practical approach to proactively mitigating wildfire risk, they could \emph{impose unnecessary difficulties if blanket outages are {conducted} throughout a power grid within a region}. In this work, we propose a method that can pinpoint the network segments with {a} higher fire ignition risk, which could help PSPS events be targeted towards those lines of the grid instead of an entire section of the grid.

\par Although several meteorological phenomena could influence wildfire ignition risk, this study focuses on high wind speed and gust conditions. This choice is mainly because the impact of wind speed on power lines' displacement can be modeled and explained by power line \textit{motion equations}, allowing for a scientific analysis of windy conditions. Additionally, from the practical point of view, PSPS events are scheduled based on extreme wind speed conditions/predictions rather than other factors. Throughout this paper, \textit{extreme} weather situations denote conditions with high wind speeds. We also limit this study to modeling conductor clashing rather than other sources of wildfire ignition inside a power system, such as phase to ground faults and contact with the surrounding vegetation, given conductor clashing is the only cause modeled in a generalized approach. Currently, no models account for other ignition causes inside a power system.

\par With the increase in the availability of meteorological sensing data and estimation techniques with high granularity, this paper presents a scientific setup for reflecting the impact of wildfire ignition risk induced by strong winds on the power system operation. To this end, we introduce a machine learning-based surrogate model that can instantly quantify the risk of wildfire ignition for each line of the grid. The surrogate model predicts and assigns \textit{scores} {that} represent the conductor-clashing probability of a power line based on real-time wind speed data. The procured quantified measure will pave the way to incorporate wildfire risk assessment into operational planning models within grids {at} the risk of wildfire. We introduce {the} Wildﬁre Risk Aware operation planning Problem (WRAP), which integrates the generated wildfire risk scores into the power system operation. The resulting model can be used to determine precisely which sections and lines of the grid should be subject to PSPS, {thereby} minimizing the area and population impacted by unnecessary outages while taking {the} potentially disastrous impacts of extreme winds into consideration.

\vspace{-0.35cm}
\subsection{Literature Review}
\par Extensive research has been dedicated to the assessment and mapping of wildfire ignition risk within different regions of the world. A fire occurrence prediction model is presented in \cite{zhang2013fire} for the northeast regions of China, where wildfire probabilities are obtained from ten variables such as altitude, slope, and distance to structures. In \cite{catry2009modeling}, eight factors are integrated into a regression model to model wildfire ignition risk in Portugal. These studies tend to predict the wildfire risk regardless of the ignition cause and focus on historical data. As a result, various overlays have been developed through the Geographic Information System (GIS). Several researchers in the field of power systems have proposed wildfire risk assessment models with the help of these tools. Authors in \cite{taylor2021framework} propose a method for the selection of power lines that should be modified to mitigate wildfire ignition risk, but no model is considered for the operation of the electricity network. A similar approach based on the spatio-temporal probability of wildfire ignition within a power system is utilized in \cite{umunnakwe2020data}. Other methods for wildfire risk assessment within power lines according to historical and geographical data are proposed in \cite{xu2016risk}, where estimation functions such as logistic regressions \cite{liu2021wildfire}, and Bayes network \cite{chen2021wildfire} are utilized. However, wildfire risk values used in these works are obtained through GIS overlays based on historical data and location, rather than ignitions induced by power lines.

\par A review of electricity grid vulnerabilities amid natural disasters, including wildfires, is presented in \cite{waseem2020electricity}. A resilient mitigation strategy for the short-term operational planning of a power system \emph{during} wildfires considering the thermal effects of wildfires on the dynamic line rating is given in \cite{teng2020enhancing}. The authors in \cite{kandanaarachchi2020early} have proposed a learning-based algorithm for early detection of wildfire ignitions due to high impedance faults. An overview of wildfire risk mitigation plans such as PSPS schedules is presented in \cite{muhs2020wildfire}, and several risk management strategies, which are carried out by utilities, are explored. An optimization framework to balance the wildfire risk and power shut-offs is presented in \cite{rhodes2020balancing}. Wildfire risk in this work is solely based on the energization status of components in the electricity grid; the impact of other factors, such as weather conditions, is not incorporated. The authors in \cite{zhou2019studies} developed a fire hazard prevention system based on meteorological factors, including  extreme wind speed. However, it is not clear how the data is modeled to forecast hazards, and no model is proposed to reflect wildfire risk in the scope of preventive measures.
The probability of wildfire ignition due to line fault and arc ignition is computed in \cite{muhs2020characterizing}, and the effect of wind is also explained. However, no model is proposed to incorporate the resulting risk values into power system operation. An overview of the recent trends in the relevant literature is displayed in Table \ref{tab:compare_wildfire}.

\setlength\tabcolsep{4pt}
\begin{table}[h!]
    \centering
    \caption{{comparison of the state of the art literature}}
    \resizebox{3.5in}{!}{\begin{tabular}{ccccc}\hline\hline
    \textbf{Reference} & \makecell{\textbf{Power} \\ \textbf{system} \\ \textbf{model}} & \makecell{ \textbf{Model} \\ \textbf{ignition} \\ \textbf{by lines}} &  \makecell{\textbf{Model} \\ \textbf{weather} \\ \textbf{conditions}} & \makecell{ \textbf{Integrated} \\ \textbf{risk} \\ \textbf{optimization}} \\\hline 
    \cite{kandanaarachchi2020early} & - & \checkmark & - & - \\ 
    \cite{taylor2021framework}, \cite{xu2016risk}, \cite{chen2021wildfire}, \cite{liu2021wildfire}  & - & - & \checkmark & - \\ 
    \cite{umunnakwe2020data}, \cite{zhou2019studies} & \checkmark & - & \checkmark & - \\ 
    \cite{teng2020enhancing} & \checkmark & \checkmark & - & - \\
    \cite{muhs2020characterizing} & \checkmark & \checkmark & \checkmark & - \\ 
    \cite{rhodes2020balancing} & \checkmark & \checkmark & - & \checkmark \\ 
    This work & \checkmark & \checkmark & \checkmark & \checkmark
    \\ \hline \hline \end{tabular}}
    \label{tab:compare_wildfire}
\end{table}

\par It is noticed in Table \ref{tab:compare_wildfire} that no work in the literature has proposed a model that incorporates the wildfire ignition risk of power lines into the system's operation planning. The only exception is reference \cite{rhodes2020balancing} where a risk model is integrated into the operation problem. However, the authors obtain \textit{relative} risk scores rather than using a physical model, and the impacts of weather conditions are neglected. {As suggested in a review article on wildfire risk management in power grids \cite{arab2021three}, there is a lack of research on wildfire prediction when power grid infrastructures could potentially cause the fire.} A gap in assessing the impact of meteorological conditions on power lines using a physical model and building a scoring mechanism to quantify the probability of conductors clashing exists in the literature. This paper aims to fill this gap by presenting a surrogate model that can efficiently quantify the risk of wildfire ignition under extreme weather conditions considering the wind speed, wind gust, wind direction, and the structure of the power lines, including  span, conductor diameter, and phase clearance. 

\vspace{-0.35cm}
\subsection{Summary of the Contributions}
The contributions of this paper are outlined as follows:
\begin{enumerate}[wide,labelwidth=!,labelindent=10pt]
\item A method for quantifying the risk of power line conductor clashing under extreme wind conditions is presented based on a physical model. We introduce a scoring method to determine the extent to which each conductor is clashing with nearby conductors. The predicted score ranges between 0 {and} 1, where 0 means no clashing is happening, while 1 means the whole conductor is under clashing. This high-granularity scoring system provides a well-quantified risk of wildfire for each individual power line. The resulting wildfire risk create{s} a scientific foundation to quantify the risk of {the} operation of each specific power line during specific extreme weather conditions.
\item A comparison of different learning algorithms for predicting the risk of wildfire is performed, and the most accurate model is proposed to predict the conductor-clashing score. The final choice is compared and verified against the nonlinear motion model, and presents a desirable performance which allows it to be confidently utilized as a surrogate model {for} the original nonlinear equations. 
\item We present WRAP to incorporate the quantified wildfire hazard risk into power system operation. No study has proposed a planning model which is able to integrate the wildfire ignition risk with power system operation. With the availability of the required meteorological data predictions, WRAP supports decision-makers {in} decid{ing} on the de-energizing of power lines for PSPS events and provides them with a mathematical foundation to study and analyze the impact of extreme weather conditions on wildfire risk.
\end{enumerate}

\vspace{-0.35cm}
\section{Quantification of Wildfire Risk Score}\label{data}
In this section, the non-linear motion of power lines due to in-plane and out-of-plane vibrations is determined to evaluate the impact of wind on the motion of power lines, which is quantified as the conductor-clashing score. A dataset is generated that considers six physical and meteorological features to find the clashing score of each line. An advantage of the power line motion equations is their generality. This model is able to accurately predict power line position under any wind condition based on physics laws, as long as all the input data are provided. Consequently, the learning task does not need to be limited to real-world data. We can learn a general surrogate model by generating enough training {and} test dataset{s} for the learning task.

\vspace{-0.35cm}
\subsection{Non-linear Equations for Motion of Power Lines}
\begin{figure}[h!] 
    \vspace{-.4cm}
    \centering
        \includegraphics[width=\columnwidth]{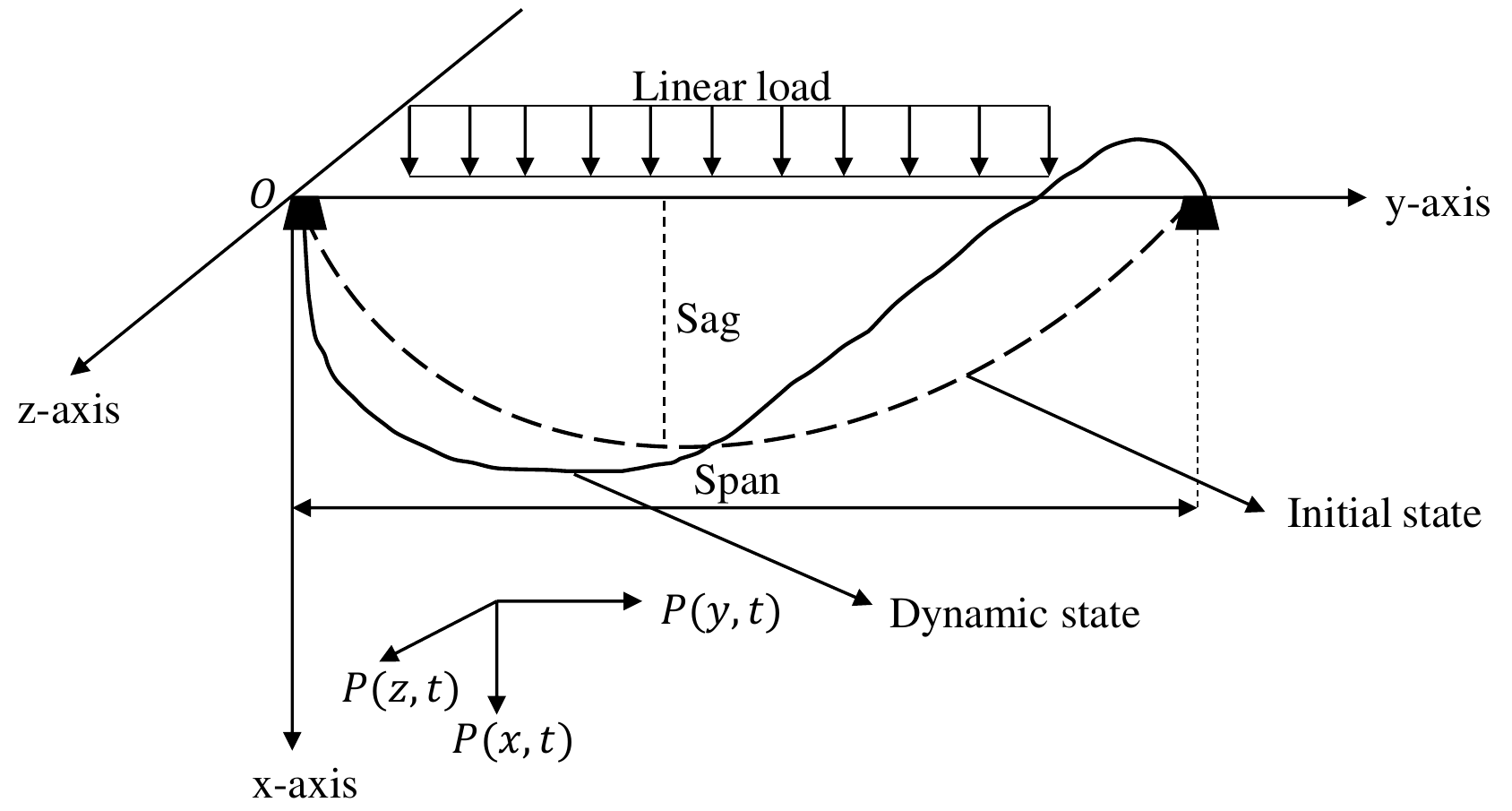}
    \vspace{-.3cm}
    \caption{Spatial configuration of the power line}
    \label{spatial_config} 
    \vspace{-.3cm}
\end{figure}

Here, the vibrational model {that} explains the movement of a power line under wind force is illustrated. Based on this model, we generated a training dataset that estimates the clashing score of a span within a power line. In-plane vibrations occur when a power line is in the plane and placed under gravity. Out-of-plane vibrations are perpendicular to the plane above. The spatial configuration of a power line is shown in Fig. \ref{spatial_config}, where $P(x,t)$, $P(y,t)$, and $P(z,t)$ correspond to the components of external excitation along the $x$, $y$, and $z$-axes, respectively.
\par The 3D geometric non-linear vibration equations of a power line are established using  Hamilton's principle \cite{zhang2018nonlinear}. Based on the boundary conditions and modeling assumptions, the 3D equations are simplified to 2D equations. Galerkin's modal truncation method converts 2D continuous Partial Differential Equations (PDEs) to discrete PDEs \cite{zhang2018nonlinear}. The discrete equations displayed in \eqref{form_mp_body} represent the coupled in-plane and out-of-plane non-linear motion of the power line. 

\begin{subequations}\label{form_mp_body}
\begin{alignat}{2}
&a_{1}\ddot{q}_{\nu}(t)+a_{2}{q}_{\nu}(t)+a_{3}{q}_{\nu}^2(t)+a_{4}{q}_{\omega}^2(t)+a_{5}{q}_{\nu}^3(t)\nonumber \\
& \hspace{3cm}+a_{6}{q}_{\omega}^2(t){q}_{\nu}(t)+a_{7}\dot{q}_{\nu}(t)=P_{\nu} \label{eq:motion_a} \\
& b_{1}\ddot{q}_{\omega}(t)+b_{2}{q}_{\omega}(t)+b_{3}{q}_{\nu}(t){q}_{\omega}(t)+b_{4}{q}_{\nu}^2(t){q}_{\omega}(t)\nonumber \\
& \hspace{3.7cm}+b_{5}{q}_{\omega}^3(t)+b_{6}\dot{q}_{\omega}(t)=P_{\omega} \label{eq:motion_b}
\end{alignat}
\end{subequations}

Here, $q_\nu$ and $q_{\omega}$ respectively denote the generalized coordinates in $x$- and $z$-axes of a power line. 
The equations for acquiring the parameters $a_1-a_7$, {$b_1-b_6$, and the initial conditions} are given in \cite{zhang2018nonlinear}.

\subsection{Wind Effect on the Motion of Power Lines}
A wind gust is a transient increase in  {the} wind speed. It occurs when there is an abrupt change from high pressure to low pressure. It is ephemeral in nature and usually lasts for 20 seconds. The discrete wind gust is used to assess the conductor{s} response to significant wind disturbances. The mathematical representation of {a} discrete wind gust is given in \eqref{wind_gust}.

\begin{equation}
% \begin{alignat}{2}
V_{wind} = \left\{\begin{matrix}
0 & & x< 0\\ 
\frac{V_m}{2} \begin{pmatrix}
1-cos(\frac{\pi x}{d_m})
\end{pmatrix} &  & 0 \leq x \leq d_m \label{wind_gust}\\ 
V_m &  & x > d_m 
\end{matrix}\right. 
% \end{alignat}
\end{equation}

Here, $V_{wind}$ is the resultant wind velocity in the power line body axis frame, $V_m$ is the wind gust amplitude, $d_m$ is the length of gust, and $x$ is the traveled distance. The gust length is the range over which the gust builds up, and the gust amplitude represents the increase in wind speed developed by the gust. The following assumptions are considered for the wind-induced vibrations of a power line.
\begin{enumerate}
    \item The wind load acts in the out-of-plane direction of the power line. This is because the effect of wind load on power line weight is negligible.
    \item The wind speed loading effect on the vibrations of power line is considered, but the aerodynamic effects are not considered. It is assumed that the wind load is considered constant and forces such as drag and lift are not modeled.
\end{enumerate}
\par The external excitation due to wind in $x$-axis direction $P(x,t)$ and $z$-axis direction $P(z,t)$ are expressed in \eqref{ext_excitation_text}.
\begin{subequations}\label{ext_excitation_text}
\begin{alignat}{2}
& P(x,t) = 0.5\rho_{a}D_{c}C_{D}(v_{a-x}-\dot{\nu}_{c})^2 \label{1b_wind_x} \\
& P(z,t) = 0.5\rho_{a}D_{c}C_{D}(v_{a-z}-\dot{\omega}_{c})^2  \label{1b_wind_z} 
\end{alignat}
\end{subequations}
Here, $\rho_{a}$ is the density of air, $D_{c}$ represents power line width facing the wind, $C_{D}$ represents power line shape factor, and $v_{a-x}$ and $v_{a-z}$ represent wind speed along the $x$ and $z$-axes, respectively. The in-plane and out-of-plane power line vibrational velocities are represented by $\dot{\nu}_{c}$ and $\dot{\omega}_{c}$, respectively. Based on the above assumptions, the expressions for $a_7$, $b_6$, $P_v$, and $P_w$ are updated considering {the} initial values of the state variables and the design hyper-parameters of the power line. The in-plane and out-of-plane responses of the power line using the Galerkin method \cite{fletcher1984computational} are represented by $\nu_c(y,t)$ and $\omega_c(y,t)$ respectively and {are} given in \eqref{galerkin_inplane_text}, where $L_c$ denotes the span of the power line.
\begin{subequations}\label{galerkin_inplane_text}
\begin{alignat}{2}
& \nu_c(y,t) = sin\begin{pmatrix}\frac{\pi y}{L_c} \end{pmatrix}\times q_\nu(t) \\
& \omega_c(y,t) = sin\begin{pmatrix}\frac{\pi y}{L_c} \end{pmatrix}\times q_{\omega}(t)
\end{alignat}
\end{subequations} 
Consequently, a new solution is obtained for \eqref{form_mp_body}, which incorporates the impact of wind speed on power line motion.

\subsection{Wildfire Risk Score Quantification}
The first step in solving \eqref{form_mp_body} is transforming it into a first-order differential equation through \eqref{state_var_1}:

\begin{equation}\label{state_var_1}
\color{black} Y_1 = {q}_{\nu}, \hspace{.71cm} Y_2 = \dot{q}_{\nu}, \hspace{.71cm} Y_3 = {q}_{\omega}, \hspace{.71cm} Y_4 = \dot{q}_{\omega}  
\end{equation}

\par Applying this transformation step turns \eqref{form_mp_body} into a system of four first-order differential equations, which are solved by the Runge-Kutta method \cite{tan2012general}. These methods are a group of numerical iterative techniques used to obtain the approximate solutions of ordinary differential equations. For example, consider the differential equation $\dot{y}=f(x,y)$ with the initial condition $y(x_0)=z_0$. If this function is discretized with a granularity of $h$, then applying the fourth-order Runge-Kutta method will result in an accumulative error in the order of $O(h^4)$.

\par To obtain the wildfire ignition risk of power lines, we first segregate each line into several segments. Solving \eqref{form_mp_body} in the presence of the wind-induced forces presented in \eqref{ext_excitation_text} determines the spatial position of each segment of a line at any given time. We suppose two horizontally (vertically) adjacent lines will clash with each other if the value of ${q}_{\omega}$ (${q}_{\nu}$) becomes greater than a certain threshold, which is set based on the distance between two neighboring conductors. Solving the motion equations for all segments of a line will determine the ratio of the line segments that are clashing. As a result, a score ranging between 0 and 1 is obtained, which determines the extent to which a power line is in contact with the neighboring conductors.

\subsection{Data Processing and Feature Importance Analysis}
Different features and meteorological conditions such as the span of the power line, conductor diameter, phase clearance, wind speed, wind gust, and wind direction angle affect the wildfire risk score. We present a surrogate machine learning model that instantly returns a high-accuracy approximation of the risk value without solving the power line motion equations for several segments of a line. In order to find the surrogate model, first a training dataset is generated according to different ranges of features based on the practical characteristics of a power system. The total number of observations in the dataset is $435,600$. Table \ref{data_samples} provides a sample of the generated dataset with scores calculated using the Runge-Kutta method. Data preprocessing comprises feature selection (dimensions of input data) and feature normalization (normalization of input data). Feature selection ensures that the power line features and meteorological conditions are contributing to predict{ing} the score.
\begin{table}[h]
    % \vspace{-0.8cm}
	\small \centering
	\caption{\color{black}Dataset Sample Generated using Runge-Kutta Method} 
	\begin{tabular}{M{0.7cm}M{1.4cm}M{0.75cm}M{0.75cm} M{1.3cm}M{1.2cm}M{0.6cm}} \hline  \hline
	    \textbf{Span} \textbf{(ft)}  & \textbf{Conductor diameter} \textbf{(mm)} & \textbf{Wind speed} \textbf{(m/s)} & \textbf{Wind gust} \textbf{(m/s)} & \textbf{Phase clearance (ft)} & \textbf{Wind direction (\textdegree)} & \textbf{Score} [0,1)  \\ \hline
		600& 33.03& 10& 12& 0.5&45  &0  \\
	    800& 34.02& 18& 16&0.5& 180   &0.01\\
% 		500& 33.03& 18& 20&0.7& 45 &  0.08\\
		1000& 31.05& 28& 30&0.5& 90  &0.12 \\
% 		400& 33.03& 26& 14&0.9&315 & 0.17 \\
% 		300& 33.03& 30& 30&0.5& 315   &0.45\\
    \hline  \hline
	\end{tabular}
	\label{data_samples}
% 	\vspace{-0.4cm}
\end{table}
\par It is necessary to quantify the importance of input features to explore the potential improvement in the input data. The input feature importance of the random forest algorithm {is} given in Table \ref{feature_importance}. For this analysis, Gini importance or Mean Decrease in Impurity (MDI) is used, which calculates the importance based on the number of times a feature is used to split a node.
It is clear that wind speed and wind gust are among the important leading features, but it is also interesting to observe the impact of span on the clashing score. Although a single important feature cannot capture the thorough relationship between the input and output, it is rational to choose the significant features to predict the conductor-clashing score. Feature normalization is helpful when features of different ranges {or} scales exist. 

\begin{table}[h]
     \vspace{-0.4cm}
   \small \centering
    \caption{\color{black} Importance of Input Features }
    \begin{tabular}{cc} \hline \hline
	\textbf{Features} & \textbf{Importance (\%)} \\ \hline
        Wind speed &  27.3\\
        Span of line &  20.9\\
        Wind gust &  13.8\\
        Wind direction & 13.5\\
        Conductor diameter & 13.0\\
        Phase clearance & 11.3\\\hline\hline
    \end{tabular}
    % \vspace{-0.4cm}
    \label{feature_importance}
\end{table}

\par It should be noted that the probability of power line conductors' clashing could be also affected by the flow of the line. This is because the heating loss of electricity increases conductor temperature, which expands line sag and escalates clashing probability. However, the results provided by \cite{kotni2014proposed} suggest that in extreme weather conditions, which are the focus of this work, heating loss is considerably mitigated by the cooling effect of wind. Hence, the impact of line current on clashing probability is neglected.

\vspace{-.1cm}
\section{Wildfire Risk Aware Problem} \label{formulation}
This section presents WRAP, an operation planning problem {that} integrates the quantified risk scores {into} power system constraints. Applying WRAP to the operation planning of a power system subject to wildfire hazard ensures that the operational constraints of the power system are satisfied. WRAP enables the power system operator to avoid energizing high-risk lines under hazardous conditions. The proposed WRAP formulation is presented in \eqref{eq:planning}, which illustrates how the procured wildfire risk quantification can be utilized in the operational planning problem of power systems.
\begin{subequations}\label{eq:planning}
\begin{alignat}{2}
&\underset{I,P_d,P_g}{\textbf{min}} \sum\limits_{t\in\mathcal{T}} \begin{Bmatrix}  \sum\limits_{l\in\mathcal{L}} {K_l} \cdot {\psi}_l^t \cdot I_l^t + \sum\limits_{i\in\mathcal{I}} {K_d} (P_{i, D}^t-P_{i, d}^t) \\
& \hspace{-2.8cm}+\sum\limits_{s\in\mathcal{S}}\sum\limits_{g\in\mathcal{G}} c_g^{s} P_{g, s}^t\end{Bmatrix} \label{eq:planning_objective}\\
& \textbf{subject to:} \nonumber\\
& \sum\limits_{s\in\mathcal{S}} P_{g, s}^t = P_g^t ,\hspace{3.5cm} \forall g\in\mathcal{G},  t\in\mathcal{T}\label{eq:planning_p_seg_sum}\\
& 0 \leq P_{g, s}^t \leq \overline{P}_{g, s} ,\hspace{2.3cm} \forall g\in\mathcal{G}, s\in\mathcal{S}, t\in\mathcal{T}  \label{eq:planning_p_seg_limit}\\
& \sum\limits_{g\in G_i} P_{g}^t + \sum\limits_{l\in LT_i} P_{l}^t= \sum\limits_{l\in LF_i} P_{l}^t + P_{i, d}^t, \hspace{.1cm} \forall i\in \mathcal{I}, t\in\mathcal{T} \label{eq:planning_powe_flow_balance_node} \\
& \underline{P}_g \leq P_{g}^t \leq \overline{P}_g,  \hspace{3.3cm} \forall g\in\mathcal{G},  t\in\mathcal{T} \label{eq:planning_power_generation_limits}\\
& \alpha P^t_{i,D}  \leq P^t_{i,d} \leq P^t_{i,D}, \hspace{2.6cm} \forall i\in \mathcal{I}, t\in\mathcal{T} \label{eq:planning_demand_served}\\
& -M(1-I_l^t)+P_l^t \leq \frac{\theta_i^t - \theta_j^t}{x_l} \leq P_l^t +M(1-I_l^t), \nonumber\\
& \hspace{5.8cm} \forall l\in \mathcal{L},  t\in\mathcal{T} \label{eq:planning_line_power_flow}\\
& -\overline{P}_l \times I_l^t \leq P_l^t \leq \overline{P}_l \times I_l^t, \hspace{1.6cm} \forall l\in \mathcal{L}, t\in\mathcal{T} \label{eq:planning_line_energized} \end{alignat}
\end{subequations}

\par The objective function is presented in \eqref{eq:planning_objective}, where the summation of the conductor clashing scores, the value of load shedding, and {the} generation cost of units is minimized. The first term in the objective penalizes the wildfire risk associated with the \emph{energized lines} by the monetization factor, $K_l$. The choice of this parameter and its implications are discussed in the case studies. Here, the quantified risk value of each line at each time is represented by ${\psi}_l^t$, where $I_l^t$ is a binary decision variable denoting the energization (if 1) or the de-energization (if 0) status of each line at each time, $l$ is the index of lines in the set of power lines $\mathcal{L}$, and $t$ is the index of time in the set of time steps $\mathcal{T}$. In the second term of the objective, the operator is penalized {by the penalty factor $K_d$} if load shedding takes place. Here, $P_{i, D}^t$ is the scheduled consumer demand, and $P_{i,d}^t$ is the amount of demand served at each bus at {a given} time, where $i$ is the index of bus{es} in set $\mathcal{I}$. The last term in the objective function calculates the generation cost of units based on a piecewise linear function approximation. The output power of each generation unit at each time is divided into different segments denoted by $P_{g,s}^t$, where $c_g^{s}$ is the generation cost, $s$ is the index of segment{s} in set $\mathcal{S}$, and $g$ is the index of generation unit{s} in set $\mathcal{G}$. The sum of power generation of each segment is equal to the total power generation of a unit{,} $P_g^t${,} as shown in \eqref{eq:planning_p_seg_sum}. The power generation of each segment is constrained to its upper and lower bounds as presented in \eqref{eq:planning_p_seg_limit}.

\par The power flow balance at each node is given in \eqref{eq:planning_powe_flow_balance_node}, where $P_l^t$ is {the} power flow of line $l$ at time $t$. The upper and lower limits of the power generation of each unit at each hour are constrained by \eqref{eq:planning_power_generation_limits}. The served demand at each bus is modeled in \eqref{eq:planning_demand_served}, where $\alpha$ is a parameter ranging from 0 to 1 and assigns the ratio of the critical demand which must be served. The relationship between the power flow of each power line with the voltage angles of its connecting buses ($\theta_i^t,\theta_j^t$) and line reactance ($x_l$) is presented with two inequalities in \eqref{eq:planning_line_power_flow}, where $M$ is an arbitrarily large number and $B_t^l,B_f^l$ are the set{s} of buses line $l$ is leading to and leaving from, respectively. The power line power flow capacity constraint is enforced in \eqref{eq:planning_line_energized}, where the thermal capacity is presented by $\bar{P}_l$. If a line is energized, i.e. $I_l^t=1$, the inequalities in \eqref{eq:planning_line_power_flow} become an equality constraint and the thermal limits in \eqref{eq:planning_line_energized} are applied. Conversely{,} if a line is de-energized (i.e.{,} $I_l^t=0$){,} the power flow in the line will be zero.

\par
\begin{figure}[h!] 
\vspace{-0.35cm}
\centering
        \includegraphics[width=3.4in]{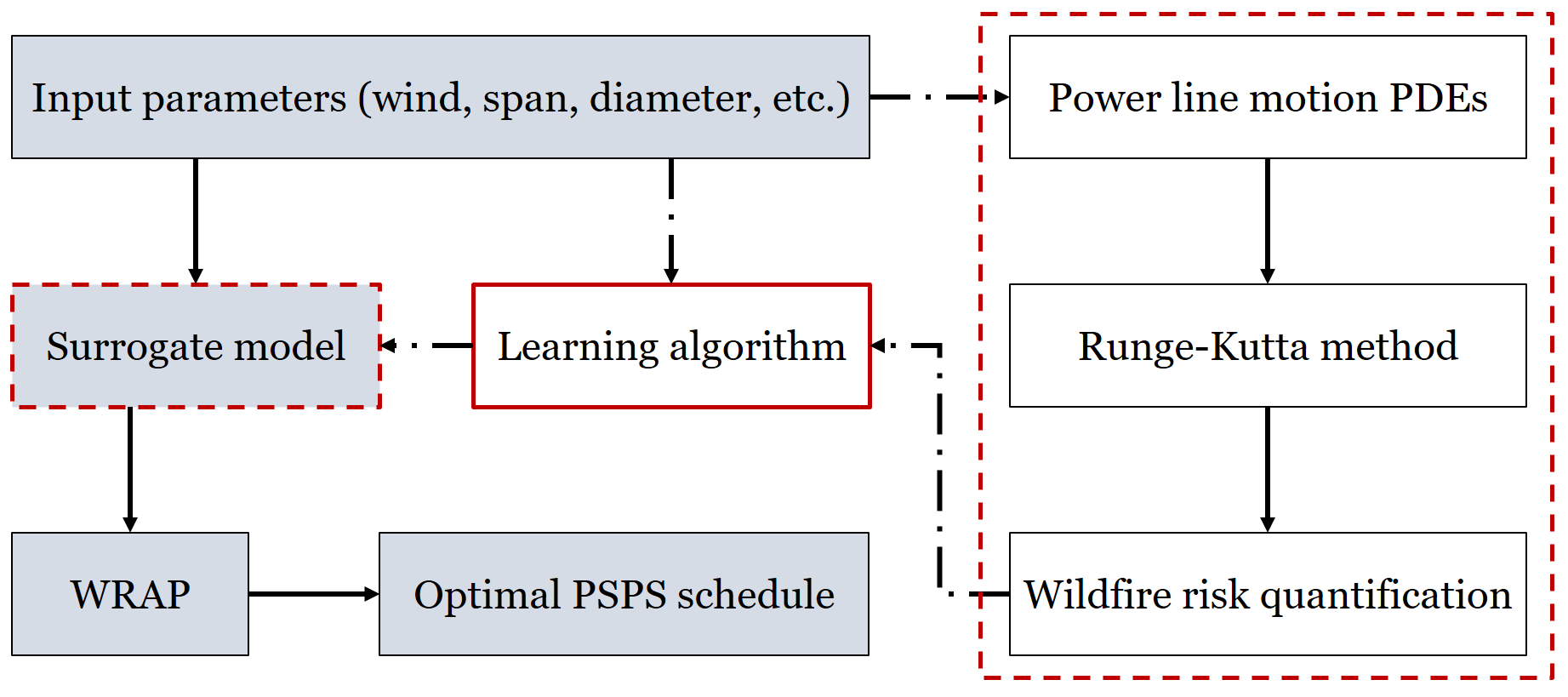}
\vspace{-0.35cm}
    \caption{ {Flowchart of the proposed wildfire risk mitigation method}}
    \label{fig:flowchart_wildfire} 
\end{figure}

 The resulting WRAP formulation in \eqref{eq:planning} is a mixed-integer linear program, which can be solved via off-the-shelf solvers. The solution to WRAP will help system operators balance the risk of wildfire in the operation of the electricity grid under the risk of wildfire. By implementing the quantified risk values of each line at each time in the WRAP, the obtained scheduling of each line's status and the output of generation units guides the system operator in decision{-}making under extreme weather conditions. Figure \ref{fig:flowchart_wildfire} illustrates the diagram of the proposed method, where the surrogate model feeds the underlying power line motion into the WRAP scheme to obtain the optimal PSPS schedule.

\vspace{-0.35cm}

% \vspace{-.35cm}
\section{Simulation Results}\label{results}
In this section, {the} performance{s} of different learning models are compared{, and} the results of the surrogate model are verified through comparison with the original motion equation model. {Next, the merits of WRAP are illustrated through comparison with two other approaches, namely naive PSPS and the method utilized by Rhodes \cite{rhodes2020balancing}. The implications of the choice of $\alpha$ for WRAP are also investigated. Finally, a discussion is presented on wildfire cost monetization along with case studies.}

\subsection{{Learning Performance}}
{In this part, several aspects of the learning process for obtaining the surrogate model are investigated.}
\subsubsection{Comparison of Various Learning Algorithms}
In order to learn a surrogate model, we performed a machine learning task using several well-known approximator structures. Here, the performance of each model is discussed and assessed with standard evaluation metrics. 
To demonstrate the merit of the proposed surrogate model, the best performing model is selected, and it is compared with the Runge-Kutta method. After careful hyper-parameter tuning, Random Forest Regression (RFR) displayed better performance than Multiple Linear Regression (MLR), {Least Absolute Shrinkage and Selection Operator (LASSO)}, Support Vector Regression (SVR), and Deep Neural Network (DNN). The optimal performance metrics for each model are shown in Table \ref{diff_learn_algo}.

\begin{table}[h!]
    \small \centering
    \caption{Comparison of Different Learning Algorithms}
    \resizebox{3.5in}{!}{\begin{tabular}{ M{2.4cm} M{0.7cm} M{1.1cm} M{0.7cm} M{1.7cm} M{1.5cm}} \hline \hline 
    \textbf{Measure} & {MLR} & {LASSO} & {SVR} & \textbf{RFR} & {DNN} \\ \hline 
    \makecell{{Mean Absolute}\\ {Error (MAE)}} & 0.03 & 0.03 & 0.02 & \textbf{0.006}\textbf{$\times10^{-3}$}&0.2$\times10^{-3}$ \\
    \makecell{{Relative MAE}} & 0.96 & 1.04 & 0.64 &  \textbf{0.16}\textbf{$\times10^{-3}$}&6.2$\times10^{-3}$  \\
    \makecell{{Root Mean Square}\\ {Error (RMSE)}} & 0.05 & 0.05 & 0.04& \textbf{0.072}\textbf{$\times10^{-3}$}&0.6$\times10^{-3}$  \\
    \makecell{{Relative RMSE}\\ {with mean}} & 1.37   & 1.52 &  1.24 & \textbf{1.96}\textbf{$\times10^{-3}$} &10$\times10^{-3}$  \\
    \makecell{{Relative RMSE}\\ {with maxmin}} & 0.11 & 0.12 & 0.10 &\textbf{0.15}\textbf{$\times10^{-3}$} &1.3$\times10^{-3}$    \\ \hline \hline 
    \end{tabular}}
    \label{diff_learn_algo}
\end{table}

\par {The} LASSO regression model is not performing better than MLR because all features are important for the learning process, and discarding a feature leads to poor performance. This is consistent with the feature importance analysis that deemed no input feature negligible. The non-linear SVR model based on {the} radial basis function kernel learns the model well and performs better than MLR and LASSO. The error evaluation metrics give the small error in {the} RFR, and this model predicts with the highest accuracy. The performance of DNN is acceptable, but it is not better than RFR. The major features {contributing} to the efficient performance of RFR are wind speed (27.3$\%$), the span of power lines (20.9$\%$), and wind gust (13.8$\%$), as shown in Table \ref{feature_importance}. These features efficiently and effectively represent the physical power line movement. Therefore, RFR can be substituted as a surrogate model to forecast the clashing score and can be incorporated into practical applications.

\subsubsection{Comparing the Surrogate Model with Nonlinear Displacement Equations - Surrogate Model Validation}
To illustrate the validity of the proposed surrogate model, a comparison between the scores predicted by the surrogate model {and} the analytical ones obtained from the physical model is presented in Table \ref{comparsion_bn_RF_PM}, where different power line spans at varying wind speeds, wind gusts, phase clearances, and wind directions are considered. It is noticed that the predicted score is almost equal to the values obtained by the Runge-Kutta (RK) model in all instances.
For example, in the second row, the score obtained by the nonlinear displacement model is 0.07142$\textbf{9}$, and the surrogate model predicted score is 0.07142$\textbf{8}$.

\par The surrogate model predicts the score with 99.99$\%$ test accuracy, where 20\% of instances are designated to the test dataset with 5-fold cross validation. Therefore, the proposed surrogate model is an accurate representation of the nonlinear conductor displacement method. Comparing the scores obtained by the surrogate model and the mathematical model validates its accuracy in several instances. By utilizing the presented surrogate model based on machine learning, risk scores are obtained instantly, and the high computation burden is no longer an issue. System operators can make informed decisions for PSPS schedule by de-energizing power lines {that have} ignition scores above a specific threshold.

\begin{table}[h!]
	\small \centering
	\caption{Comparison Between the Predictions of Physical and Surrogate Models} 
	\begin{tabular}{ M{0.6cm} M{0.6cm} M{0.6cm} M{0.6cm} M{2.1cm} M{0.5cm} M{0.8cm} M{0.8cm }} \hline  \hline
	     \textbf{$L_{c}$} \textbf{(ft)}& \textbf{$d$} \textbf{(mm)} & \textbf{$ V_\omega$} \textbf{(m/s)} & \textbf{$ V_g$} \textbf{(m/s)} & \textbf{Phase clearance (ft)} & \textbf{Angle (\textdegree)} &  \multicolumn{2}{c}{\textbf{Score}}\\ \hline
		 & & & & & & \textbf{RFR}&\textbf{RK}   \\\hline
% 		 600& 33.03& 10& 12& 0.5&45  & \textbf{0}&0  \\
		 800& 33.03& 26& 26&1.5& 45 & \textbf{0.0714}  &0.0714\\% (RFR=0.0714, RK=0.0714)
		 500& 33.03& 18& 20&0.7& 45 & \textbf{0.0857} & 0.0857\\ % (RFR=0.0857, RK=0.085)
		%4 & 900& 33.03& 24& 16&1.1& 315 &  &0.02 \\\hline
		 1000& 33.03& 30& 28&0.7& 315 & \textbf{0.1285} &0.1285 \\% (RFR=0.1285, RK=0.128)
		 400& 33.03& 26& 14&0.9&315 & \textbf{0.1714}&0.1714 \\ % (RFR=0.171428, RK=0.171)
		 300& 33.03& 30& 30&0.5& 315 & \textbf{0.4571}  &0.4571\\\hline  \hline% (RFR=0.45714, RK=0.457)
	\end{tabular}
	\label{comparsion_bn_RF_PM}
\end{table}

\subsubsection{Impact of Various Meteorological and Structural Features on the Conductor Clashing Score}
\begin{figure}[h!] 
% \vspace{-0.3cm}
\centering
        \includegraphics[width=.9\columnwidth]{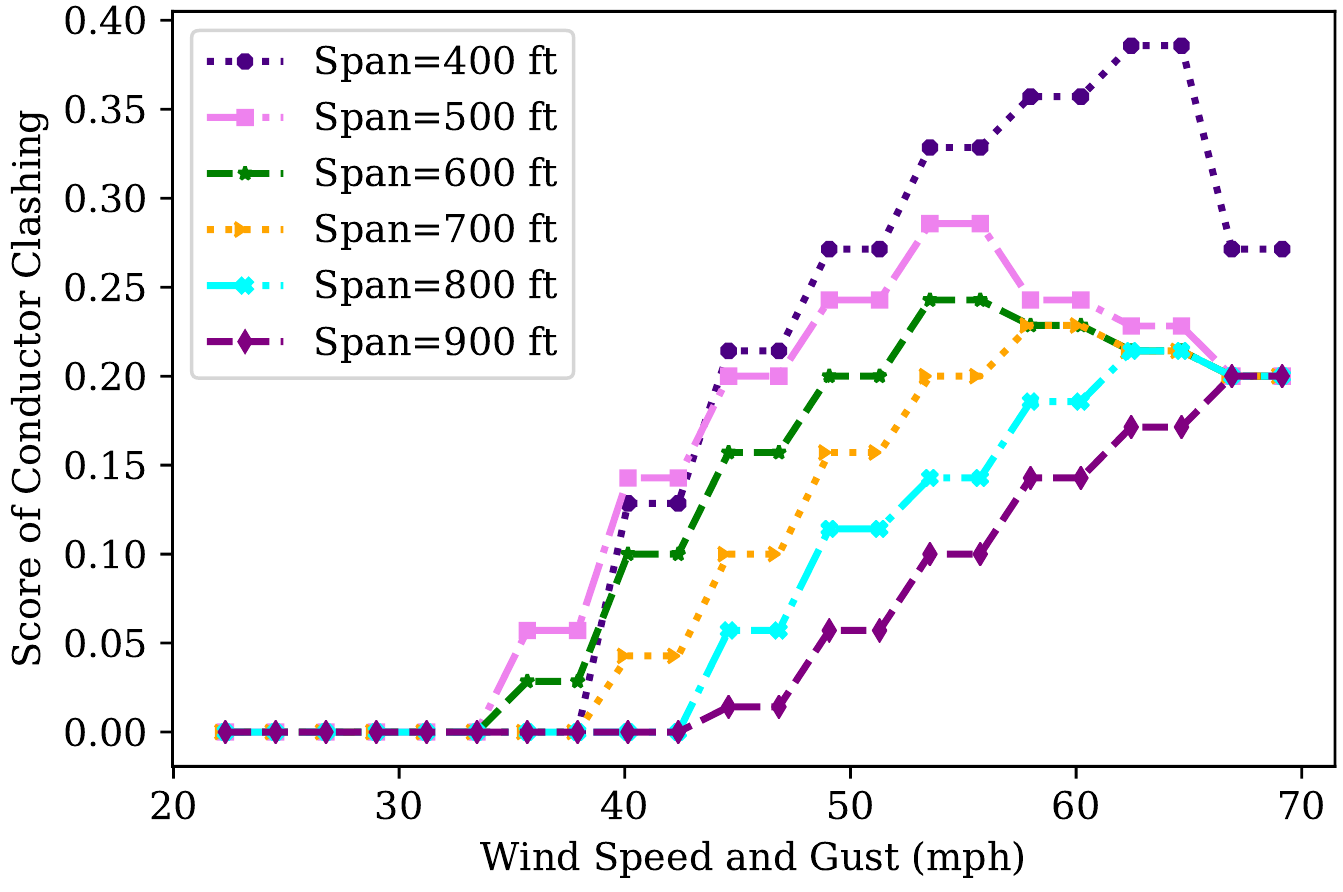}
    \vspace{-0.35cm}
    \caption{Conductor clashing score for different spans using the surrogate model at d = 33.03 mm, phase clearance = 0.5 ft, and angle = 45\textdegree}
    \label{diff_spans} 
    % \vspace{-0.2cm}
\end{figure}

In this part, the surrogate model predicts the clashing score by varying different input features. The conductor clashing scores for different line spans and wind speeds are shown in Fig. \ref{diff_spans}, where it is illustrated that the clashing scores will be different for various line spans at the same wind speed. Here, the conductor diameter, phase clearance, and angle are kept constant to investigate the effect of power line span and wind speed on the conductor clashing. The increase in wind speed and wind gust leads to an increase in the conductor clashing score. The conductors with a larger span have a lower conductor clashing score for the same wind speed and wind gust because a larger span means conductors are heavier and harder to move. It is noticed that the conductor clashing scores of lines with shorter spans drop with the increase in wind speed above a certain threshold. This is because lighter conductors of various phases will move together at higher wind speeds, reducing  conductor clashing. Thus, simultaneous consideration of meteorological and structural features is necessary. 

\begin{figure}[h!] 
% \vspace{-0.3cm}
    \centering
        \includegraphics[width=.9\columnwidth]{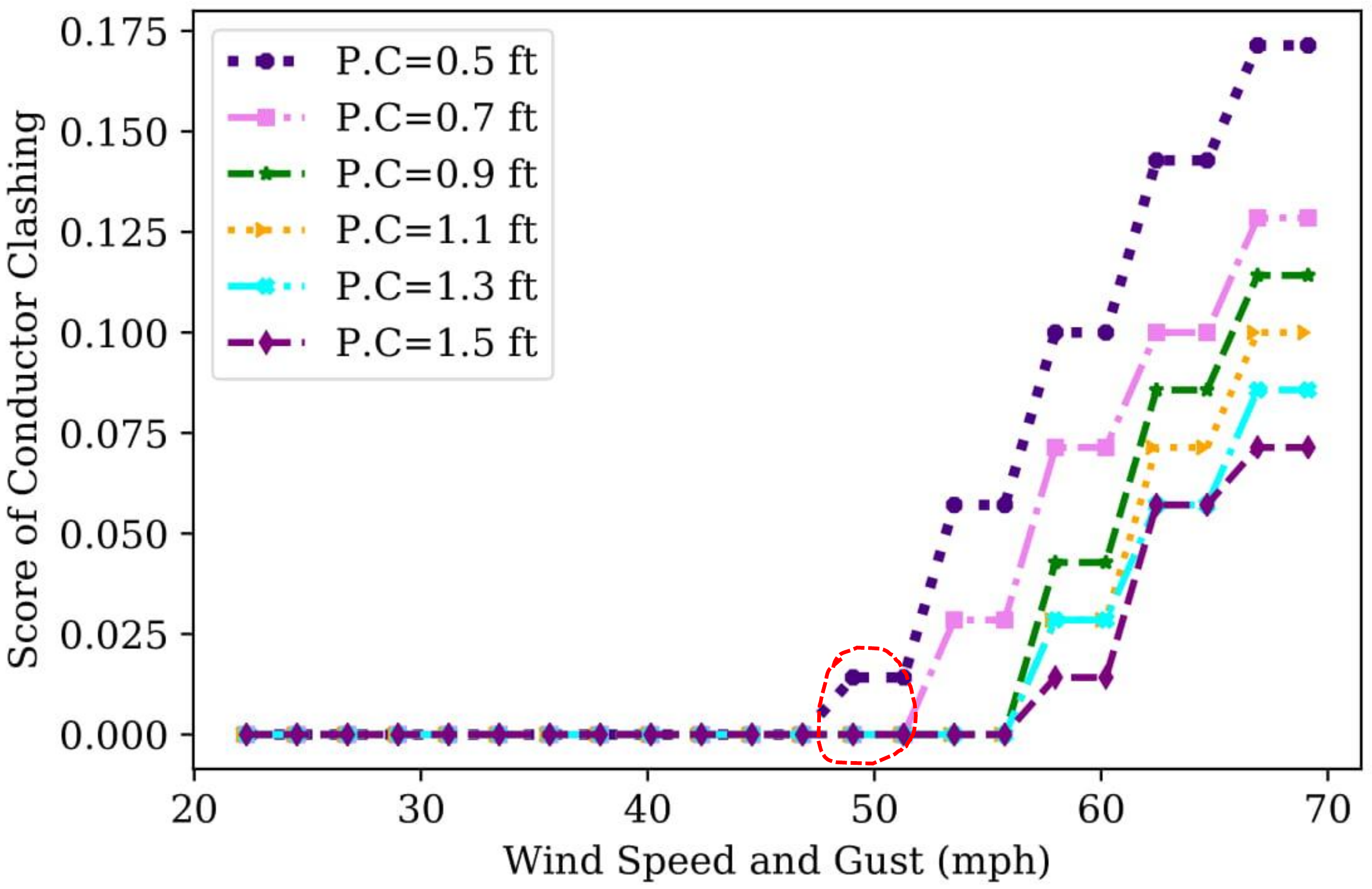}
        \vspace{-0.35cm}
    \caption{Conductor clashing score for different phase clearances using the surrogate model at span = 1000 ft, d = 33.03 mm, and angle = 45\textdegree}
    \label{diff_clc} 
    % \vspace{-0.2cm}
\end{figure}

The conductor clashing scores for various phase clearances (P.C) are shown in Fig. \ref{diff_clc}, where the span, diameter, and angle are set to 1000 ft, 33.03 mm, and 45$^{\circ}$, respectively. A smaller phase clearance {results in} a higher clashing score. Based on this observation, it is concluded that choosing a one-size-ﬁts-all threshold for wind speed (e.g., 50 mph) for line de-energizing is not a reliable approach. 

When wind speed is constant, the conductor clashing score decreases with increasing phase clearances since. With a large phase clearance, the conductors move further apart, and smaller sections of the conductors will slap each other.  
% \vspace{-0.2cm}

\subsection{WRAP Performance and Comparison Metrics}
In this part, the performance of the proposed wildfire-resilient operation framework is investigated. First, WRAP is applied to find the optimal operation schedule of a test 6-bus network. The effectiveness of WRAP in mitigating the wildfire risk of a network under extreme weather conditions is illustrated. We then present two other cases to compare the risk mitigation plans obtained by WRAP with the naive PSPS scenario as well as the method proposed by Rhodes et al. in \cite{rhodes2020balancing}. {The wildfire risk is reduced by de-energizing more power lines, which increases the amount of load shedding.}
All of the simulations are performed using the Julia \cite{bezanson2017julia} programming language, where optimizations are handled with the JuMP \cite{DunningHuchetteLubin2017} module. The average simulation run time for the 6-bus and 30-bus networks is 0.04 second and 0.66 second, respectively.

\par
\begin{figure}[h!] 
    \vspace{-1.15cm}
    \centering
        \includegraphics[width=\columnwidth]{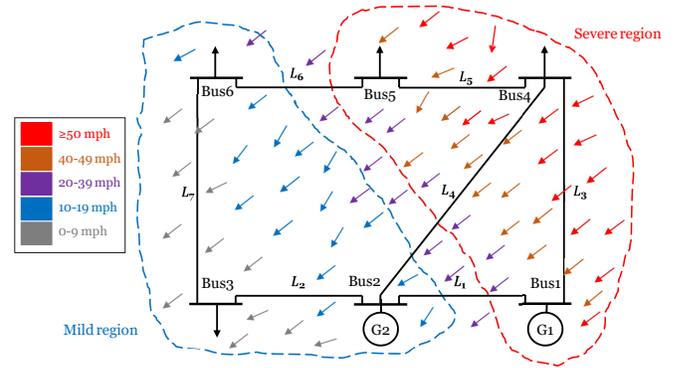}
    \vspace{-1.65cm}
    \caption{Demonstration of wind impact on a 6-bus system}
    \label{6bus_network}
    \vspace{-0.25cm}
\end{figure}

 Implementing WRAP allows system operators to find the optimal schedule for the de-energizing of lines, given the wildfire risk and the penalty {parameters}. Figure \ref{6bus_network} displays the diagram of the test 6-bus grid that is subject to wind. This network has {seven} lines and {two} generation units. The wind direction and speed are represented by colored arrows, where the respective wind speed range {for} each color is shown in the legend. It is observed that this network can be regarded as two regions with severe and mild wind conditions. The wildfire risk score for the lines placed in the mild regions was 0 throughout the whole day. However, the lines in the severe region have a non-zero wildfire risk during most periods of the day. The wildfire risk of these lines changes according to a hypothetical wind curve and only reaches zero during hours 8-17 of the day.

\par
\begin{figure}[h!] 
    \centering
        \includegraphics[width=3.5in]{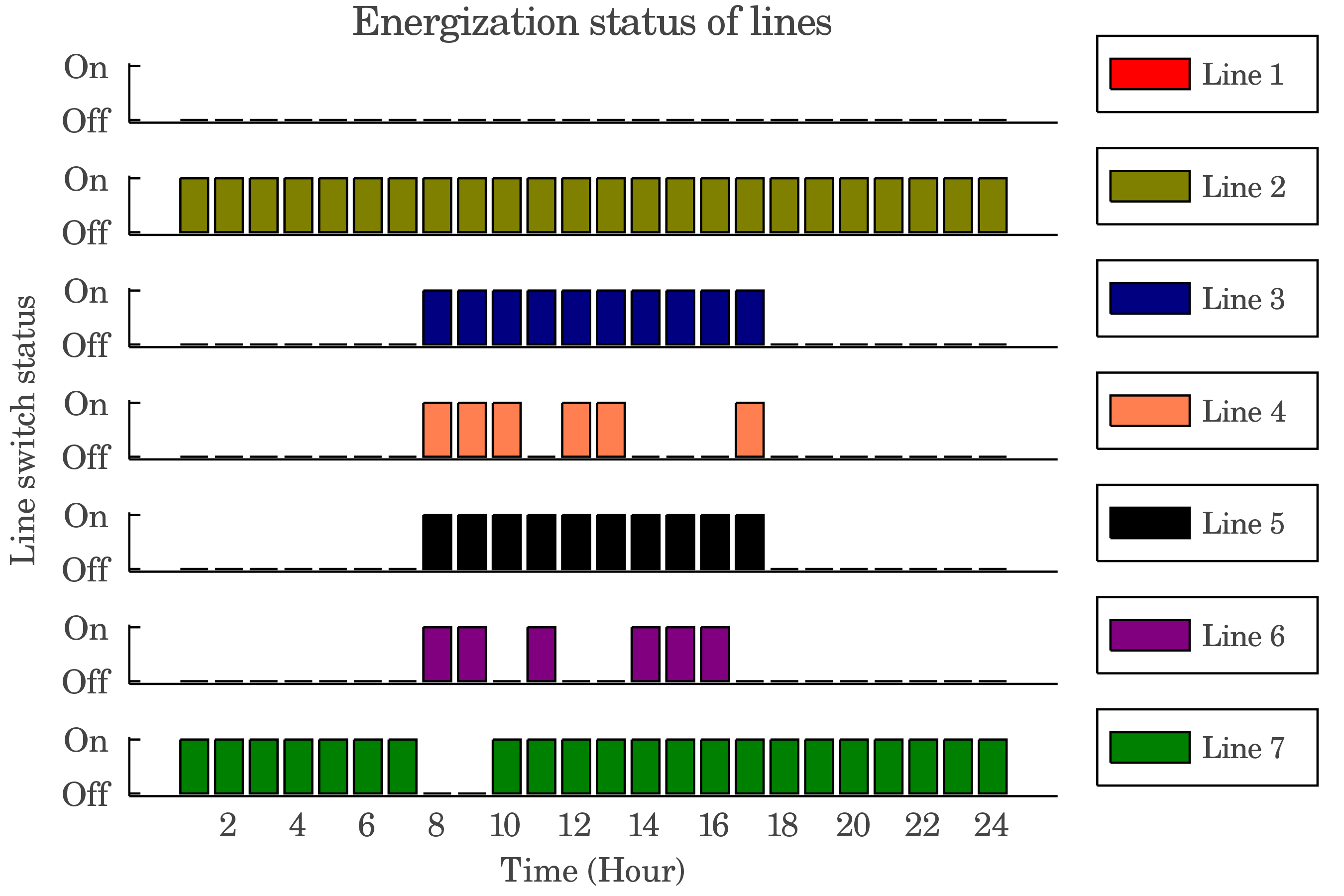}
    \vspace{-.35cm}
    \caption{{Line switching status obtained by WRAP}}
    \label{fig:wrap_I_line}
\end{figure}

 The presented surrogate model {provides} system operators with a tool to decide on the de-energization of lines based on the quantified risk scores, load continuity priorities{, and expected wildfire costs}. The line switching schedules as a result of utilizing WRAP for the 24 hour operation of the test 6-bus network are presented in Fig. \ref{fig:wrap_I_line}. It is noticed that line 1 is kept de-energized throughout all periods of the day while line 2 is energized for the whole day. This is because the wildfire risk associated with line 2 is equal to zero at all periods of the day. As a matter of fact, the total daily wildfire risk for the operation schedule of this network obtained by WRAP is equal to 0. Since wildfire risk is associated with the conductor clashing probability of power lines, this means the lines are only energized when their risk is 0.

\par It is noticed that compared with a naive PSPS scenario and the approach proposed by Rhodes et al. \cite{rhodes2020balancing}, WRAP results in considerably lower operation costs. One reason is that WRAP incurs no expected wildfire costs since wildfire risk is maintained at zero in its operations. Additionally, compared with the naive PSPS scenario and the Rhodes approach, WRAP serves the highest amount of demand to consumers. A summary of the results obtained by each of these methods is presented in Table \ref{tab:compare_wrap}. In the naive PSPS approach, utilities shut off power to segments of the network where wildfire ignition risk is deemed high. To simulate the naive PSPS scenario, it was assumed that lines in the extreme wind region are switched off. As a result, the naive PSPS case considers the operation of the network with a fixed configuration where lines 3, 4, and 5 are disconnected, and all of the other lines are kept switched on for the whole duration of the day. %The operation costs of the mentioned naive PSPS scenario add up to \$74.4M.

\setlength\tabcolsep{3pt} 
\begin{table}[h!] \centering \caption{{Comparing performance of WRAP with two other approaches}} \label{tab:compare_wrap}
\begin{tabular}{cccc} \hline \hline
Method  & Naive PSPS& Rhodes et al. \cite{rhodes2020balancing} & WRAP\\  \hline
Total wildfire risk & 2.32 & 0 & 0 \\
Total served demand (MWh) & 2398.56 & 1917.8  & \textbf{2760.2} \\
Expected wildfire costs ($\times1000\$$) & 72,696  & 0 & 0 \\
Percentage of load shedding (\%) & 37.47 & 50.0  & \textbf{28.04}  \\
Total operation costs ($\times1000\$$) & 73,434  & 974.4 & \textbf{561.4}\\   \hline \hline
\end{tabular}
\end{table}

\par
\begin{figure}[h!] 
    \centering
        \includegraphics[width=3in]{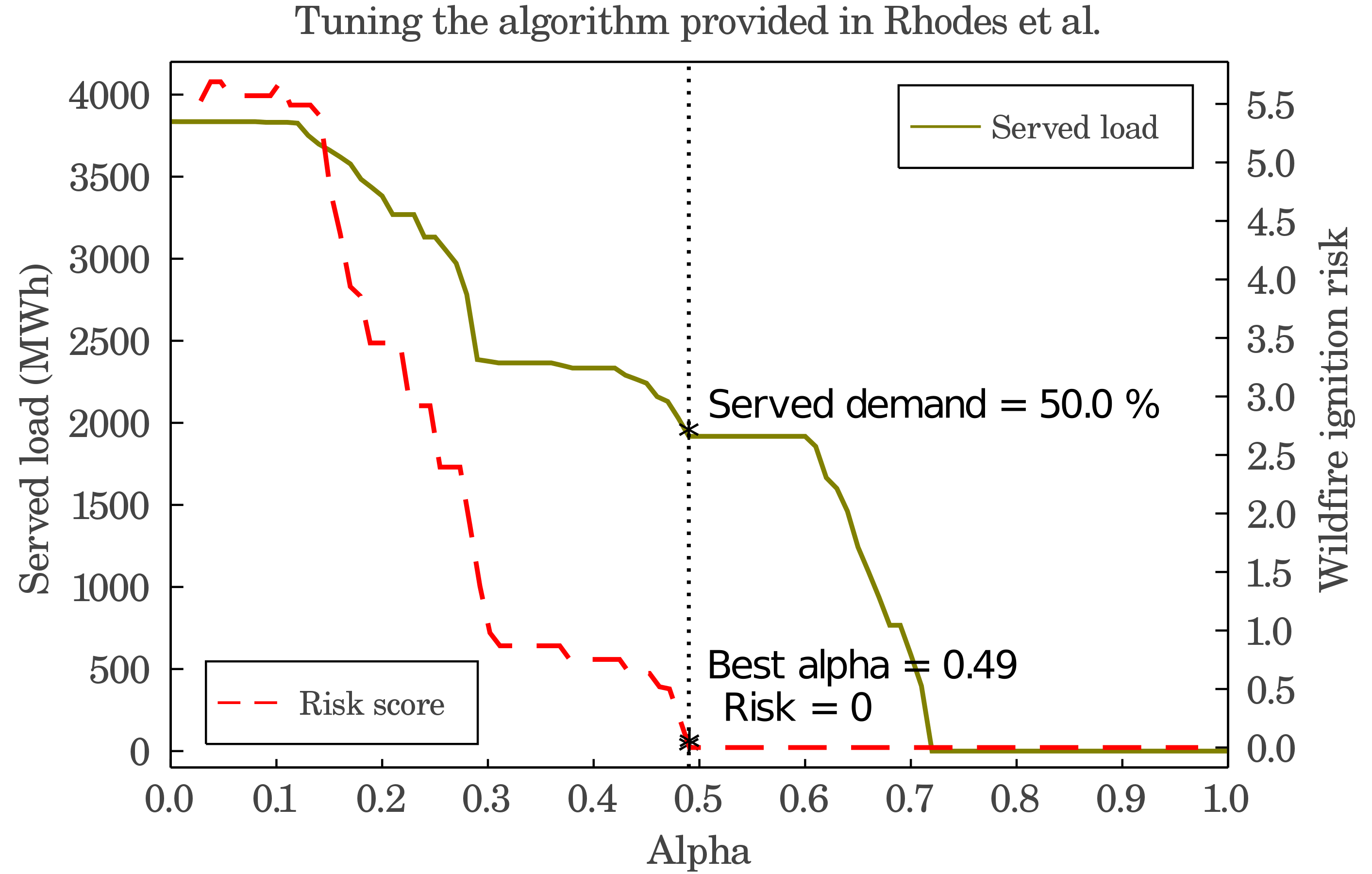}
    \vspace{-.35cm}
    \caption{{Tuning the control parameter in Rhodes et al. \cite{rhodes2020balancing}}}
    \label{fig:rhodes_tuning}
\end{figure}

 In addition to a naive PSPS scenario, the performance of WRAP is also compared with the approach introduced in \cite{rhodes2020balancing}. Rhodes et al. propose a method that integrates the wildfire ignition risk associated with each element of the power system into the optimal operation problem of the network. The model in \cite{rhodes2020balancing} returns a relative value for the wildfire risk and requires tuning of the control parameter `Alpha`. This value adjusts the weight of load serving and wildfire risk in the objective function. The tuning process for this algorithm is displayed in Fig. \ref{fig:rhodes_tuning}. With the best tuning, this method reduces the wildfire risk to zero. However, it fails to meet 50.0\% of the daily demand as opposed to the proposed WRAP method, which only misses 28.04\% of the demand. The approach presented in \cite{rhodes2020balancing} yields a more conservative mitigation plan than the one obtained by WRAP, because it does not account for the implications of weather conditions on wildfire ignition risk.

\par
\begin{figure}[h!] 
\vspace{-0.3cm}
    \centering
    \centering{\includegraphics[width=.9\columnwidth]{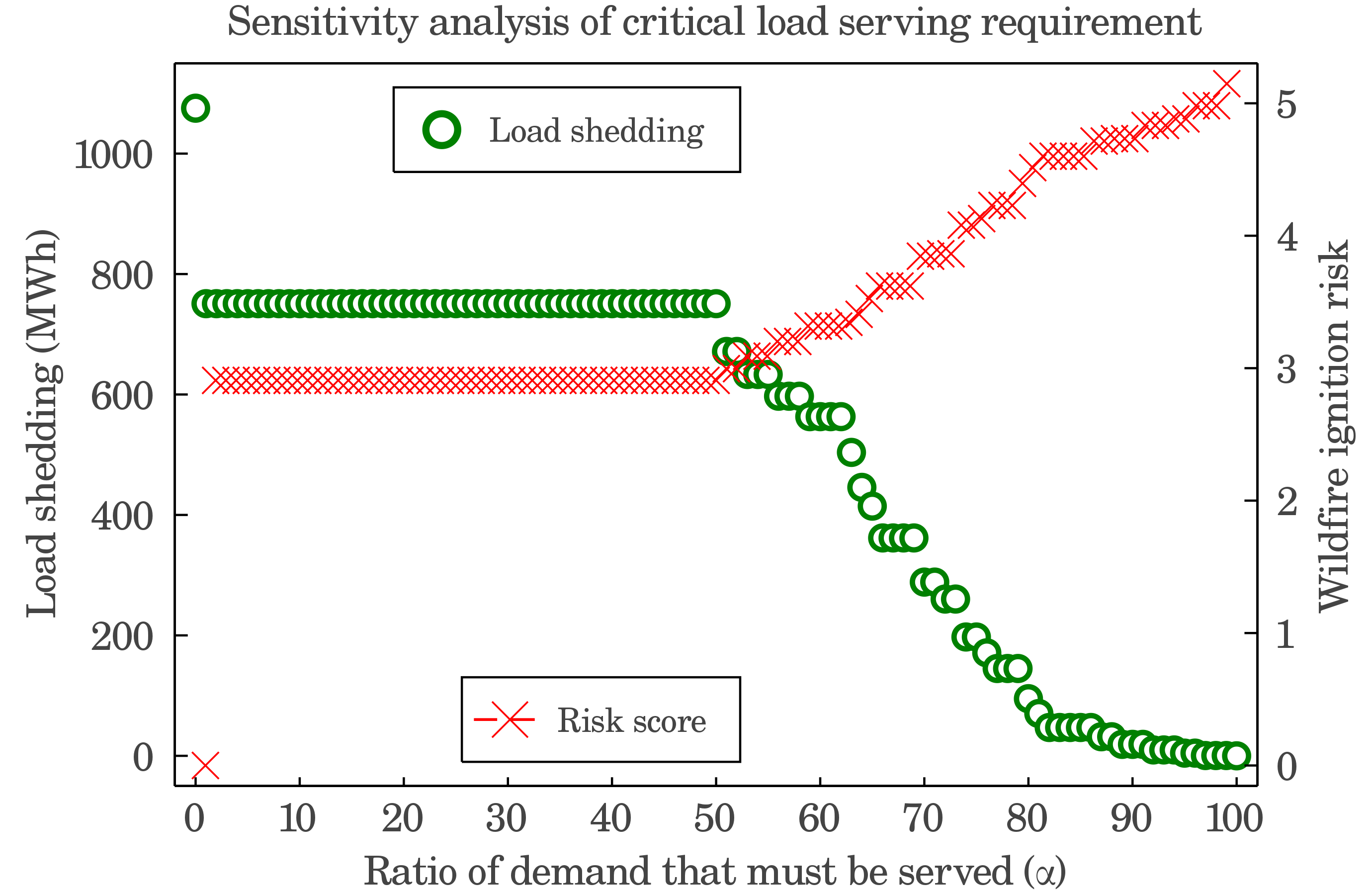}}
\vspace{-0.3cm}
    \caption{{The balance between load shedding and cumulative fire hazard risk based on enforced load serving in the 6-bus system}}
    \label{fig:shed_tolerance}
\vspace{-0.7cm}
\end{figure}

 An analysis of the trade-off between load shedding and fire hazard risk by varying the load-serving percentage ($\alpha$) is presented in Fig. \ref{fig:shed_tolerance}. With the gradual change in the value of $\alpha$ from 0 to 100\%, load shedding decreases from {1075.4} MWh to 0 MWh, and {the} cumulative fire hazard risk increases from 0 to {5.15}. The increase in the load-serving percentage leads to an increase in objective cost from {\$0.56M} to {\$230.8M}, which signifies the inherent cost of fire hazard risk. It is interesting to observe that by increasing {the} load-serving percentage from {1\% to 50\%}, the load shedding and cumulative score do not change. This is because the lines that were energized to serve the {1\%} of demand can still support {50\%} of demand without needing additional support. Therefore, the fire hazard risk taken into account to serve the load remains unchanged. {Another interesting observation in this figure is the jump in risk and load shedding values by raising $\alpha$ from 0 to 1\%. When the load serving requirement is set to 0, the system is able to serve 2760.2 MWh of demand without scheduling lines that bear wildfire risk, which results in zero cumulative wildfire risk. However, even if 1\% of demand is forced to be served at each location and time interval, the system has to include lines with wildfire ignition risk in its operation.}

\subsection{Demonstration of WRAP on the IEEE 30-Bus System}
\par
\begin{figure}[h!] 
\vspace{-1.2cm}
    \centering
    \includegraphics[width=\columnwidth]{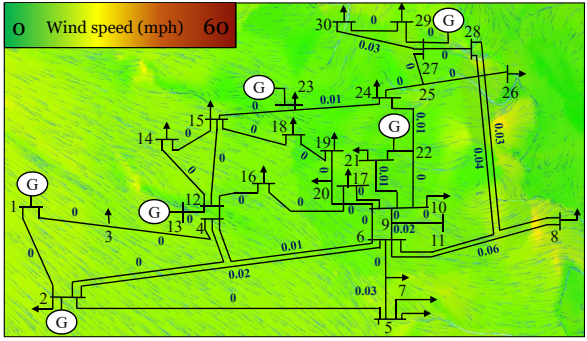}
    \vspace{-1.65cm}
    \caption{The quantified risk of wildfire ignition for low wind speed case in the IEEE 30-bus system during 1$^{\text{st}}$ hour of the day}
    \label{risk_30_bus_low_wind} 
\vspace{-0.3cm}
\end{figure}

 In this section, the presented surrogate model is applied to quantify the risk of wildfire ignition in {the} modified IEEE 30-bus system during low and high wind speed situations. The network is overlaid on a similar geographical area where the meteorological information can be accessed at \cite{web:sdge}. The available meteorological data includes wind speed, wind gust, and wind direction. Other structural features, including the span of line, conductor diameter, and phase clearance, are based on the data of the IEEE 30-bus system. The wildfire ignition score during {the} low wind speed scenario is shown in Fig. \ref{risk_30_bus_low_wind}. Here, the wildfire ignition score ranges between 0 {and} 0.06, with the majority of them being 0 due to low wind speed, with the maximum wind speed being 24 mph (11 m/s). The highest risk value of 0.06 belongs to the line connecting buses 6 and 8 because this region has the highest wind speed and gust of 24 mph, causing a relatively higher but still small fire hazard risk.

\par
\begin{figure}[h!] 
 \vspace{-1.2cm}
   \centering
        \includegraphics[width=\columnwidth]{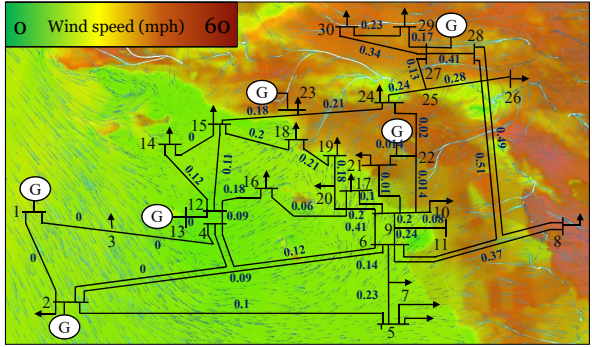}
\vspace{-1.65cm}
    \caption{The quantified risk of wildfire ignition for high wind speed case in the IEEE 30-bus system during 1$^{\text{st}}$ hour of the day}
    \label{risk_30_bus_high_wind} 
    \vspace{-0.3cm}
\end{figure}

 The wildfire ignition score during a high wind speed scenario is shown in Fig. \ref{risk_30_bus_high_wind}. Here, the maximum wind speed reaches 62 mph (28 m/s), and the fire risk score ranges between 0 and 0.51, where the highest score is associated with the highest wind speed. The aggravating effect of higher wind speeds on ignition risk is noticeable in these two cases. Although the peak wind speed here is only 2.5 times that in the low wind speed scenario, the maximum score in the high wind speed scenario is 8.5 times that obtained in the low-speed scenario. The system operator may adjust the value of $\alpha$ to balance load-serving and the risk value. The presented WRAP model provides a tool for the system operator to balance fire hazard risk and service continuity {by} leveraging the rendered risk score.

\par
\begin{figure}[h!] 
    \vspace{-0.2cm}
    \centering
        \includegraphics[width=3.5in]{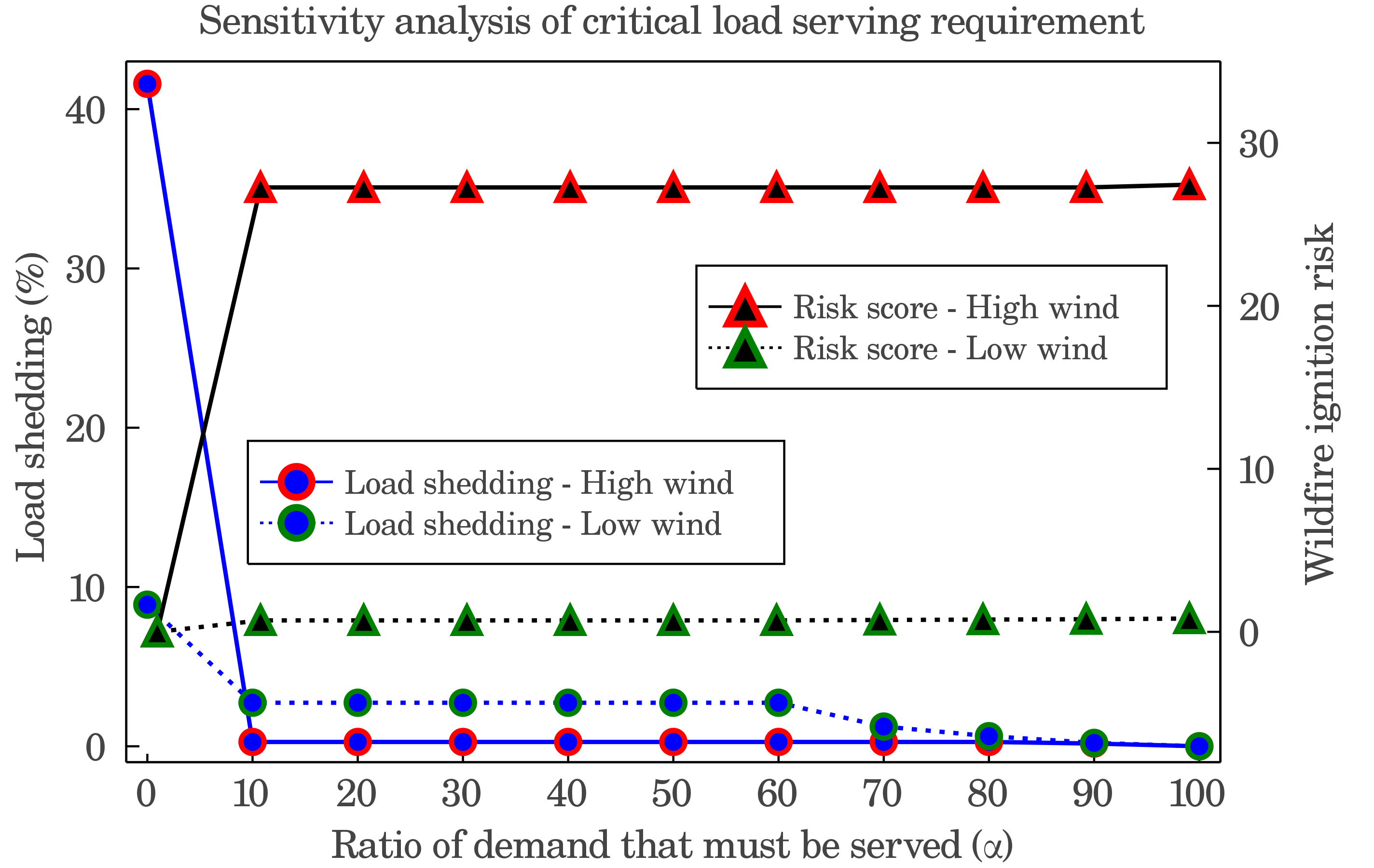}
    \vspace{-0.35cm}
    \caption{{The balance between load shedding and cumulative fire hazard risk based on enforced load serving in the IEEE 30-bus system}}
    \label{low_high_wind_30_bus} 
\vspace{-0.2cm}
\end{figure}

 The balance between fire hazard risk and load shedding based on the percentage of forced load-serving is presented in Fig. \ref{low_high_wind_30_bus}. In the \textit{low wind} scenario, at 0\% enforced load serving, the objective cost is {\$193,024} with {a} load shedding of {8.9\%} and the cumulative fire hazard risk of 0. As the value of $\alpha$ is increased, the amount of load shedding is decreased while the {wildfire} risk is elevated. It is observed that with raising $\alpha$ from 0\% to 10\%, load shedding drops to {2.7\%} while the {wildfires} risk is increased to {0.7}. With 100\% enforced load serving, the cumulative fire hazard risk is only {0.81} for the 24 hours of operation.

\par The system operator's challenge to balance service continuity with load-serving is more complicated during the \textit{high wind} speed scenario, as shown in Fig. \ref{low_high_wind_30_bus}, where the objective cost is {\$788,187} and load shedding is {41.6\%} at 0\% load serving. The reason behind the difference in the percentage of load shedding in these two scenarios is the difference in the number of lines with a risk score of 0. By increasing the enforced load-serving from 10\% to {80\%}, similar to what was observed in the 6-bus system case, the load shedding and cumulative fire hazard risk remain constant. It is noticed that the fire hazard risk in high wind scenarios sharply escalates when the system is forced to serve any amount of demand, while no further sudden surges are observed as the percentage of the served load is increased, and the load shedding is decreased. This can be associated with the network structure, where to serve an even low{er} percentage of the load, several lines must become energized. With the 100\% enforced load serving, the objective cost is increased to {\$593.2M}, and the fire hazard score is increased to {27.44} for the 24 hours of operation. {A summary of the observations in these cases is presented in Table \ref{tab:compare_low_high}.}

\begin{table}[h!] \centering \caption{{Comparing the results of WRAP in two different wind conditions}} \label{tab:compare_low_high} \centering
\begin{tabular}{cccc} \hline \hline
$\alpha$ (\%)  & 0 & 50 & 100 \\  \hline
\textbf{Low wind condition} & &  &  \\ 
Total wildfire risk & 0 & 0.701 & 0.814 \\
Expected wildfire costs ($\times1000\$$) & 0 & 6,289.4 & 6,629.9 \\
Percentage of served demand (\%) & 91.11 & 97.27  & 100.0  \\
Total operation costs ($\times1000\$$) & 193.0  & 6,373.3 & 6,664.1 \\ \hline
\textbf{High wind condition} & & &  \\ 
Total wildfire risk & 0.006 & 27.264 & 27.435 \\
Expected wildfire costs ($\times1000\$$) & 12.9  & 590,385.8 & 593,162.0 \\
Percentage of served demand (\%) & 58.41 & 99.72  & 100.0  \\
Total operation costs ($\times1000\$$) & 78.8  & 590,429.4 & 593,200.8 \\ \hline \hline
\end{tabular}
\end{table}

\subsection{Wildfire Risk Monetization}
The first term in the objective function \eqref{eq:planning_objective} aims to monetize the overall wildfire risks. According to \cite{dale2009true}, the unpredictable nature of wildfire aftermaths makes monetizing the risk of wildfire very challenging. Based on metrics including but not limited to suppression cost, property loss, damage to facilities, rehabilitation costs, the authors in \cite{dale2009true} performed a cost analysis on several wildfire instances. The reported statistics suggest that a wildfire could cost between \$400-\$22,500 per acre. To monetize the risk of wildfires, we incorporated $K_L$ into our model, a parameter that varies based on the conditions of each region. The monetization values considered for the 33-bus test system are illustrated in Fig. \ref{fig:30bus_cost}. Regions with more facilities and human activity are considered to have higher costs.

\begin{figure}[h!] 
    \vspace{-1.3cm}
    \centering
        \includegraphics[width=3.5in]{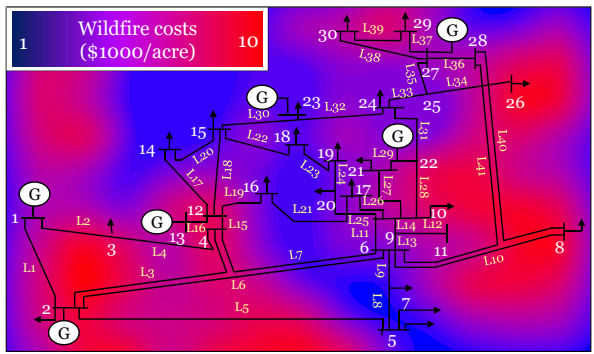}
    \vspace{-1.35cm}
    \caption{{The wildfire risk monetization factor in the IEEE 30-bus system}}
    \label{fig:30bus_cost} 
        \vspace{-0.3cm}
\end{figure}

\par Our simulations indicate that WRAP tends to utilize lines that are associated with lower wildfire costs. Results of a case study on the 30-bus system with $\alpha=50\%$ are presented in Fig. \ref{fig:30bus_cost}. This scatter graph shows the correlation between the wildfire monetization factor of each line and its utilization. It is noticed that lines with higher wildfire costs are not scheduled by WRAP as much as potentially cheaper lines. The size of the filled circles in this figure represents the total amount of risk associated with the daily operations of each line. It can be observed that larger circles are placed in the lower part of this plot. This observation suggests that implementing WRAP limits the operation schedule of lines with higher ignition risk scores.

\begin{figure}[h!] 
    \vspace{-.1cm}
    \centering
        \includegraphics[width=3.5in]{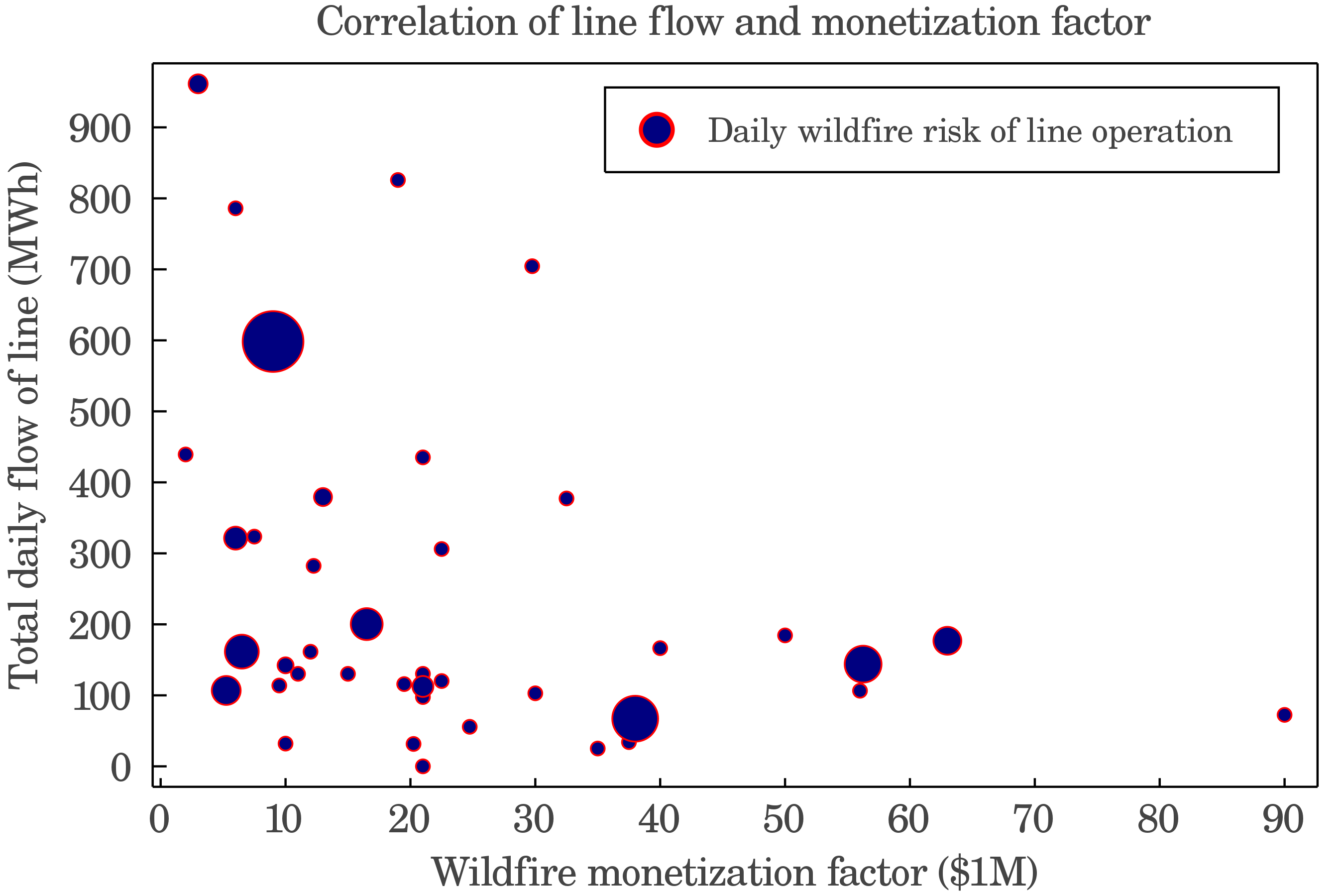}
    \vspace{-.35cm}
    \caption{{Correlation of wildfire monetization factor and daily utilization of line}}
    \label{fig:30bus_cost} 
        \vspace{-0.4cm}
\end{figure}

\section{Conclusions} \label{conclusions}
In this paper, a machine-learning-based surrogate model is developed that quantifies the risk of wildfire ignition by power lines under different weather conditions with a very high accuracy. We also introduced wildfire risk aware operation planning problem, an optimization framework that enables system operators to schedule the switching of power lines based on the quantified risk values to balance service continuity and wildfire risk. With the availability of geographical data ,this tool can be applied to instantly pinpoint the power lines that need to be de-energized by system operators during \emph{public safety power shut-off} events while allowing flexible decision-making. The results of our method show that compared with the naive public safety power shut-off scenario, the wildfire risk aware operation planning problem schedule reduces wildfire risk and operation costs of the system while increasing the amount of load served. Compared with another approach in the literature, the proposed model reduces operation costs by 42\% without increasing the wildfire risk. Several analyses on the control parameter $\alpha$, which determines the amount of critical load, are also performed. A discussion on the wildfire monetization risk factor indicates that wildfire risk aware operation planning problem tends to prioritize utilization of lines with lower costs and limit the operation of high risk lines. Developing a model that is able to incorporate all of the factors that affect the risk of wildfire ignition within a power system such as temperature, humidity, vegetation, topological, and historical data is considered as the extension of this study.

\ifCLASSOPTIONcaptionsoff
\fi
\bibliographystyle{IEEEtran}
\bibliography{quantifying_final.bbl}

% Generated by IEEEtran.bst, version: 1.14 (2015/08/26)
\begin{thebibliography}{10}
\providecommand{\url}[1]{#1}
\csname url@samestyle\endcsname
\providecommand{\newblock}{\relax}
\providecommand{\bibinfo}[2]{#2}
\providecommand{\BIBentrySTDinterwordspacing}{\spaceskip=0pt\relax}
\providecommand{\BIBentryALTinterwordstretchfactor}{4}
\providecommand{\BIBentryALTinterwordspacing}{\spaceskip=\fontdimen2\font plus
\BIBentryALTinterwordstretchfactor\fontdimen3\font minus
  \fontdimen4\font\relax}
\providecommand{\BIBforeignlanguage}[2]{{%
\expandafter\ifx\csname l@#1\endcsname\relax
\typeout{** WARNING: IEEEtran.bst: No hyphenation pattern has been}%
\typeout{** loaded for the language `#1'. Using the pattern for}%
\typeout{** the default language instead.}%
\else
\language=\csname l@#1\endcsname
\fi
#2}}
\providecommand{\BIBdecl}{\relax}
\BIBdecl

\bibitem{jolly2015climate}
W.~M. Jolly, M.~A. Cochrane, P.~H. Freeborn, Z.~A. Holden, T.~J. Brown, G.~J.
  Williamson, and D.~M. Bowman, ``Climate-induced variations in global wildfire
  danger from 1979 to 2013,'' \emph{Nature Communications}, vol.~6, no.~1, pp.
  1--11, 2015.

\bibitem{weber2020spatiotemporal}
K.~T. Weber and R.~Yadav, ``Spatiotemporal trends in wildfires across the
  western united states (1950--2019),'' \emph{Remote Sensing}, vol.~12, no.~18,
  p. 2959, 2020.

\bibitem{syphard2015location}
A.~D. Syphard and J.~E. Keeley, ``Location, timing and extent of wildfire vary
  by cause of ignition,'' \emph{International Journal of Wildland Fire},
  vol.~24, no.~1, pp. 37--47, 2015.

\bibitem{pge_report}
{Pacific Gas and Electric Company}, ``Pacific {G}as and {E}lectric {C}ompany
  amended 2019 wildfire safety plan,'' \emph{Available Online:
  \url{https://bit.ly/36d30Si}}, 2019.

\bibitem{mitchell2013power}
J.~W. Mitchell, ``Power line failures and catastrophic wildfires under extreme
  weather conditions,'' \emph{Engineering Failure Analysis}, vol.~35, pp.
  726--735, 2013.

\bibitem{psps}
{California Public Utilities Commission}, ``Utility public safety power shutoff
  plans (de-energization),'' \emph{Available Online:
  \url{https://www.cpuc.ca.gov/psps/}}, 2020.

\bibitem{kandanaarachchi2020early}
S.~Kandanaarachchi, N.~Anantharama, and M.~A. Munoz, ``Early detection of
  vegetation ignition due to powerline faults,'' \emph{IEEE Transactions on
  Power Delivery}, 2020.

\bibitem{sotolongo2020}
\BIBentryALTinterwordspacing
M.~Sotolongo, C.~Bolon, and S.~H. Baker, ``California power shutoffs:
  Deficiencies in data and reporting,'' in \emph{Initiative for energy
  justice}, 2020. [Online]. Available: \url{https://bit.ly/3wfPibX}
\BIBentrySTDinterwordspacing

\bibitem{gentle2012concurrent}
J.~P. Gentle, ``Concurrent wind cooling in power transmission lines,'' Idaho
  National Laboratory (INL), Tech. Rep., 2012.

\bibitem{lihong2006parameters}
L.~Lihong, H.~Yi, L.~Jinglu, and Z.~Xiangjun, ``Parameters for wind caused
  overhead transmission line swing and fault,'' in \emph{TENCON 2006- IEEE
  Region 10 Conference}.\hskip 1em plus 0.5em minus 0.4em\relax IEEE, 2006, pp.
  1--4.

\bibitem{sutlovic2019analysis}
E.~Sutlovic, I.~Ramljak, and M.~Majstrovic, ``Analysis of conductor clashing
  experiments,'' \emph{Electrical Engineering}, vol. 101, no.~2, pp. 467--476,
  2019.

\bibitem{jazebi2019review}
S.~Jazebi, F.~De~Leon, and A.~Nelson, ``Review of wildfire management
  techniques—part {I}: Causes, prevention, detection, suppression, and data
  analytics,'' \emph{IEEE Transactions on Power Delivery}, vol.~35, no.~1, pp.
  430--439, 2019.

\bibitem{murphy2021}
\BIBentryALTinterwordspacing
P.~Murphy, ``Preventing wildfires with power outages: the growing impacts of
  {C}alifornia’s public safety power shutoffs,'' in \emph{Energy and
  Environment Program}, 2021. [Online]. Available: \url{https://bit.ly/3hD26DX}
\BIBentrySTDinterwordspacing

\bibitem{cnbc2019}
\BIBentryALTinterwordspacing
P.~Stevens, ``{PG\&E} power outage could cost the {C}alifornia economy more
  than \$2 billion.''\hskip 1em plus 0.5em minus 0.4em\relax CNBC, 2021.
  [Online]. Available: \url{https://cnb.cx/36e21kR}
\BIBentrySTDinterwordspacing

\bibitem{fox2019}
\BIBentryALTinterwordspacing
M.~Philips, ``Oxygen-dependent {C}alifornia man dies 12 minutes after {PG\&E}
  cuts power to his home.''\hskip 1em plus 0.5em minus 0.4em\relax Fox News,
  2021. [Online]. Available: \url{https://fxn.ws/3hlKcGK}
\BIBentrySTDinterwordspacing

\bibitem{zhang2013fire}
H.~Zhang, X.~Han, and S.~Dai, ``Fire occurrence probability mapping of
  northeast china with binary logistic regression model,'' \emph{IEEE Journal
  of Selected Topics in Applied Earth Observations and Remote Sensing}, vol.~6,
  no.~1, pp. 121--127, 2013.

\bibitem{catry2009modeling}
F.~X. Catry, F.~C. Rego, F.~L. Ba{\c{c}}{\~a}o, and F.~Moreira, ``Modeling and
  mapping wildfire ignition risk in portugal,'' \emph{International Journal of
  Wildland Fire}, vol.~18, no.~8, pp. 921--931, 2009.

\bibitem{taylor2021framework}
S.~Taylor and L.~A. Roald, ``A framework for risk assessment and optimal line
  upgrade selection to mitigate wildfire risk,'' \emph{arXiv preprint
  arXiv:2110.07348}, 2021.

\bibitem{umunnakwe2020data}
A.~Umunnakwe, M.~Parvania, H.~Nguyen, J.~D. Horel, and K.~R. Davis,
  ``Data-driven spatio-temporal analysis of wildfire risk to power systems
  operation,'' \emph{IET Generation, Transmission \& Distribution}, 2022.

\bibitem{xu2016risk}
K.~Xu, X.~Zhang, Z.~Chen, W.~Wu, and T.~Li, ``Risk assessment for wildfire
  occurrence in high-voltage power line corridors by using remote-sensing
  techniques: a case study in hubei province, china,'' \emph{International
  journal of remote sensing}, vol.~37, no.~20, pp. 4818--4837, 2016.

\bibitem{liu2021wildfire}
H.~Liu, E.~Zhou, L.~Fan, Z.~Rao, W.~Chen, and Y.~Zhou, ``Wildfire risk
  assessment of transmission-line corridors based on logistic regression,'' in
  \emph{2021 IEEE 4th International Electrical and Energy Conference
  (CIEEC)}.\hskip 1em plus 0.5em minus 0.4em\relax IEEE, 2021, pp. 1--5.

\bibitem{chen2021wildfire}
W.~Chen, Y.~Zhou, E.~Zhou, Z.~Xiang, W.~Zhou, and J.~Lu, ``Wildfire risk
  assessment of transmission-line corridors based on na{\"\i}ve bayes network
  and remote sensing data,'' \emph{Sensors}, vol.~21, no.~2, p. 634, 2021.

\bibitem{waseem2020electricity}
M.~Waseem and S.~D. Manshadi, ``Electricity grid resilience amid various
  natural disasters: Challenges and solutions,'' \emph{The Electricity
  Journal}, vol.~33, no.~10, p. 106864, 2020.

\bibitem{teng2020enhancing}
F.~Teng, ``Enhancing power distribution grid resilience against massive
  wildfires,'' Ph.D. dissertation, The George Washington University, 2020.

\bibitem{muhs2020wildfire}
J.~W. Muhs, M.~Parvania, and M.~Shahidehpour, ``Wildfire risk mitigation: A
  paradigm shift in power systems planning and operation,'' \emph{IEEE Open
  Access Journal of Power and Energy}, vol.~7, pp. 366--375, 2020.

\bibitem{rhodes2020balancing}
N.~Rhodes, L.~Ntaimo, and L.~Roald, ``Balancing wildfire risk and power outages
  through optimized power shut-offs,'' \emph{arXiv preprint arXiv:2004.07156},
  2020.

\bibitem{zhou2019studies}
T.~Zhou, B.~Li, C.~Wu, Y.~Tan, L.~Mao, and W.~Wu, ``Studies on big data mining
  techniques in wildfire prevention for power system,'' in \emph{2019 IEEE 3rd
  Conference on Energy Internet and Energy System Integration (EI2)}.\hskip 1em
  plus 0.5em minus 0.4em\relax IEEE, 2019, pp. 866--871.

\bibitem{muhs2020characterizing}
J.~Muhs, M.~Parvania, H.~T. Nguyen, and J.~A. Palmer, ``Characterizing
  probability of wildfire ignition caused by power distribution lines,''
  \emph{IEEE Transactions on Power Delivery}, 2020.

\bibitem{arab2021three}
A.~Arab, A.~Khodaei, R.~Eskandarpour, M.~P. Thompson, and Y.~Wei, ``Three lines
  of defense for wildfire risk management in electric power grids: A review,''
  \emph{IEEE Access}, 2021.

\bibitem{zhang2018nonlinear}
M.~Zhang, G.~Zhao, and J.~Li, ``Nonlinear dynamic analysis of high-voltage
  overhead transmission lines,'' \emph{Shock and Vibration}, 2018.

\bibitem{fletcher1984computational}
C.~A. Fletcher, ``Computational {G}alerkin {M}ethods.''\hskip 1em plus 0.5em
  minus 0.4em\relax Springer, 1984.

\bibitem{tan2012general}
D.~Tan and Z.~Chen, ``On a general formula of fourth order {R}unge-{K}utta
  method,'' \emph{Journal of Mathematical Science \& Mathematics Education},
  vol.~7, no.~2, pp. 1--10, 2012.

\bibitem{kotni2014proposed}
L.~Kotni, ``A proposed algorithm for an overhead transmission line conductor
  temperature rise calculation,'' \emph{International Transactions on
  Electrical Energy Systems}, vol.~24, no.~4, pp. 578--596, 2014.

\bibitem{bezanson2017julia}
J.~Bezanson, A.~Edelman, S.~Karpinski, and V.~B. Shah, ``Julia: A fresh
  approach to numerical computing,'' \emph{SIAM review}, vol.~59, no.~1, pp.
  65--98, 2017.

\bibitem{DunningHuchetteLubin2017}
I.~Dunning, J.~Huchette, and M.~Lubin, ``Jump: A modeling language for
  mathematical optimization,'' \emph{SIAM Review}, vol.~59, no.~2, pp.
  295--320, 2017.

\bibitem{web:sdge}
\BIBentryALTinterwordspacing
``{SDGE} {W}eather {A}wareness {S}ystem,'' last accessed 6 July 2021. [Online].
  Available: \url{https://sdgeweather.com}
\BIBentrySTDinterwordspacing

\bibitem{dale2009true}
L.~Dale, \emph{The true cost of wildfire in the Western US}.\hskip 1em plus
  0.5em minus 0.4em\relax Western Forestry Leadership Coalition, 2009.

\end{thebibliography}

\begin{IEEEbiography}[{\includegraphics[width=1in,height=1.25in,clip,keepaspectratio]{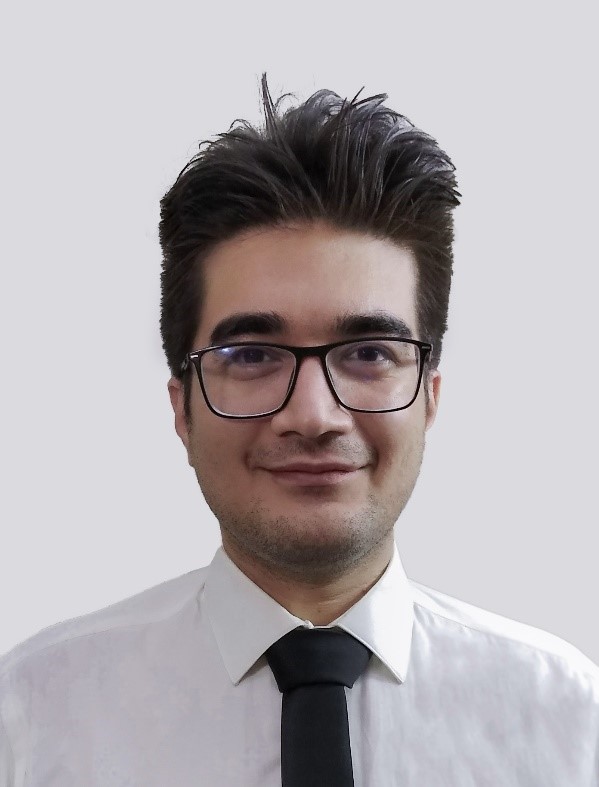}}]%
{Reza Bayani} (S’20) received the B.Sc. degree from Sharif University of Technology, Tehran, Iran, in 2013 and the M.Sc. degree from Iran University of Science and Technology, Tehran, Iran, in 2016, both in Electrical Engineering. He is currently pursuing his Ph.D. degree in Electrical and Computer Engineering at the University of California San Diego, La Jolla, USA, and San Diego State University, San Diego, USA. His current research interests include power system operation and planning, machine learning, and transportation electrification. He is a Ph.D. fellow of California State University’s Chancellor’s Doctoral Incentive Program and was awarded SDSU’s 2022-2023 university graduate fellowship.
\end{IEEEbiography}

\begin{IEEEbiography}[{\includegraphics[width=1in,height=1.25in,clip,keepaspectratio]{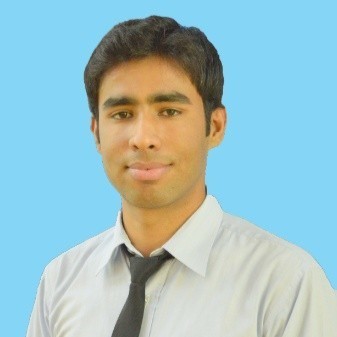}}]%
{Muhammad Waseem} received the B.S. degree in electrical engineering from the School of Electrical and Computer Engineering, University of Engineering and Technology, Lahore, Pakistan, in 2017. He is currently pursuing the M.S. degree in electrical engineering from the Department of Electrical Engineering, San Diego State University, San Diego, CA, USA. His current research interests include smart grid, optimization, and power system operation and planning.
\end{IEEEbiography}

\begin{IEEEbiography}[{\includegraphics[width=1in,height=1.25in,clip,keepaspectratio]{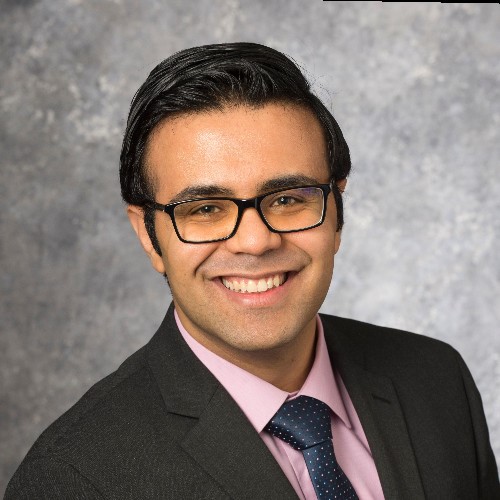}}]%
{Saeed D. Manshadi} (M’18) is an Assistant Professor with the Department of Electrical and Computer Engineering at San Diego State University. He was a postdoctoral fellow at the Department of Electrical and Computer Engineering at the University of California, Riverside. He received the Ph.D. degree from Southern Methodist University, Dallas, TX; the M.S. degree from the University at Buffalo, the State University of New York (SUNY), Buffalo, NY and the B.S. degree from the University of Tehran, Tehran, Iran all in electrical engineering. He serves as an editor for IEEE Transactions on Vehicular Technology. His current research interests include smart grid, transportation electrification, microgrids, integrating renewable and distributed resources, and power system operation and planning.
\end{IEEEbiography}

\begin{IEEEbiography}[{\includegraphics[width=1in,height=1.25in,clip,keepaspectratio]{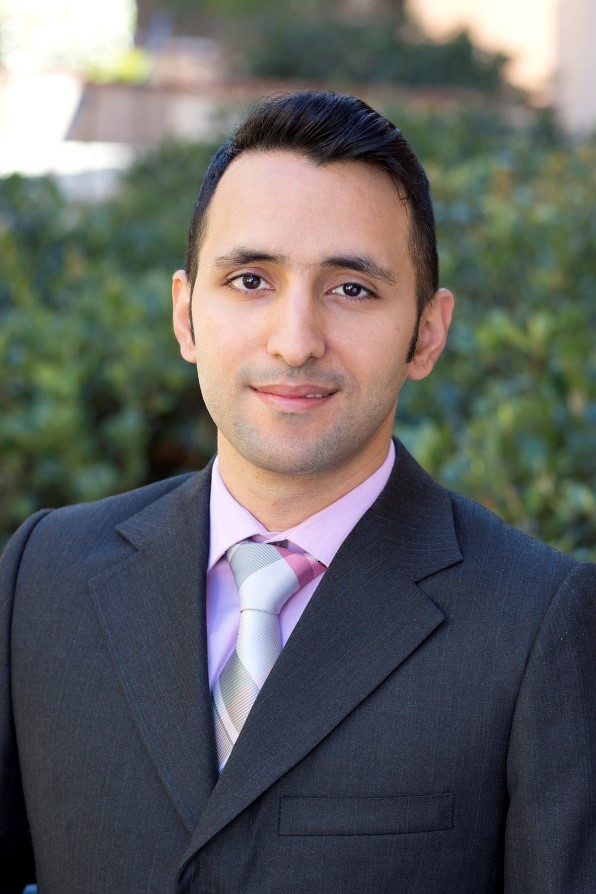}}]%
{Hassan Davani} is an Assistant Professor at the SDSU Department of Civil, Construction and Environmental Engineering, specialized in water resources and climate change forecasting. All his experience in academia and industry has focused on application of advanced computational techniques to explore emerging climatic stressors and flooding challenges on the civil infrastructure. He has over 20 peer-reviewed journal publications to date, and has collaborated internationally between the US and Europe. He serves as a member for two committees of the American Society of Civil Engineers: International Participation Committee, and Urban Water Resources Council.
\end{IEEEbiography}

\end{document}